\documentclass[aps,prb,twocolumn,floatfix,showpacs]{revtex4}
 \pdfoutput=1
\usepackage{graphicx} 
\usepackage{calc} 
\usepackage{amsfonts}  
\usepackage{amsmath} 
\usepackage{feynmf}
\usepackage{sparticles}

\def\kvec{ {\bf k}}

\def\qvec{{\bf q}}

\def\ktil{\bar{k}}
\def\qtil{\bar{q}}

\def\0til{\bar {0}}

\def\sumslashD{\mathop{\sum \kern-1.4em - \kern 0.5em}}

\def\sumscrossD{\mathop{\sum \kern-1.5em \times \kern 0.5em}}

\def\sumslash{\mathop{\sum \kern-1.2em -\kern 0.5em}}
\def\sumscross{\mathop{\sum \kern-1.2em \times \kern 0.5em}}
\def\intslash{\mathop{\int \kern-0.9em -\kern 0.5em}}
\def\intslashD{\mathop{\int \kern-1.1em -\kern 0.5em}}

\def\kperp{ {k_{\perp}}}
\def\qperp{ {q_{\perp}}}

\usepackage{psfrag}
\begin{document}
\title{Extended quantum criticality  of  low-dimensional  superconductors near a spin-density-wave instability}        
 
\author{A. Sedeki$^1$, D. Bergeron$^1$ and C. Bourbonnais$^{1,2}$}
\affiliation{$^1$Regroupement Qu\'ebecois sur les Mat\'eriaux de Pointe,
  D\'epartement de physique, Universit\'e de Sherbrooke, Sherbrooke,
  Qu\'ebec, Canada, J1K-2R1 }
\affiliation{$^2$Canadian Institute for Advanced Research, Toronto, Canada}
 
\date{\today}

\begin{abstract}
 We use the  renormalization group method  to study  normal state properties  of  quasi-one-dimensional superconductors nearby a   spin-density-wave instability. On the basis of  one-loop  scattering amplitudes for the quasi-one-dimensional electron gas, the integration of the renormalization group equations  for the two-loop single particle Matsubara  self-energy leads to   a nonFermi-liquid   temperature downturn of the momentum-resolved quasi-particle weight  over   most part of the Fermi surface. The amplitude of the downturn correlates with    the entire instability line for superconductivity, defining an extended quantum critical region  of the phase diagram  as a function of   nesting deviations of the Fermi surface.  One     also extracts  the downward renormalization of interchain hopping amplitudes    at  arbitrary low temperature in  the normal phase. By  means of   analytical continuation of the Matsubara self-energy,  one-particle spectral functions are obtained with respect to  both energy and temperature  and  their  anomalous  features  analyzed in connection with the sequence of instability lines of the phase diagram. The quasi-particle scattering rate is found to develop  an unusual temperature dependence, which is  best  described  by the superimposition of a    linear   and  quadratic    $T$ dependences. The nonFermi-liquid  linear-$T$ component correlates with the temperature scale $T_c$ of the superconducting instability over an extended range of nesting deviations,  whereas  its  anisotropy  along  the Fermi surface is predicted to parallel  the momentum profile of a   d-wave pairing gap on the Fermi surface. We examine the implications of our results for low dimensional unconventional superconductors, 
   in particular the Bechgaard salts series of quasi-one-dimensional organic conductors,
  but also the pnictide and cuprate superconductors where several common features are observed.
  \end{abstract}
\maketitle 

\section{Introduction}
The metallic state of  unconventional superconductors  reveals  most often   anomalies  when   
  superconductivity takes place in the  vicinity of a density-wave  instability. This is  particularly    exemplified in the  (TMTSF)$_2X$  series of quasi-one-dimensional organic  conductors, the so-called  Bechgaard salts, which develop superconductivity (SC) on the verge of a spin-density-wave (SDW) state  under pressure\cite{Jerome80,Jerome82,Bourbon08,Brown08,Bourbon11}. Anomalous enhancement of the NMR spin-lattice relaxation rate, $T_1^{-1}$,\cite{Brown08,Wu05,Wzietek93,CreuzetRMN,Kimura11} by SDW   fluctuations,  and   linear   temperature resistivity,\cite{DoironLeyraud09a,DoironLeyraud09,DoironLeyraud10}  stand out as among   the most distinctive features shown    by  the  metallic phase of the  Bechgaard salts over a broad range of pressure.     Even more  striking  is the apparent correlation   these anomalies   display     with  the size of the SC critical temperature, $T_c$.\cite{DoironLeyraud09a,DoironLeyraud09,Taillefer10}   
    
 In the case of electrical transport,  the correlation of $T_c$ with the linear-$T$ resistivity term shows how   quasi-particle scattering   on low energy SDW fluctuations is  related   to the  strength of Cooper pairing and  hence to  the  mechanism    causing superconductivity\cite{Taillefer10,DoironLeyraud09,DoironLeyraud10}.  This parallel change   exhibited over the whole pressure range where $T_c$ is finite,   contrasts  with the usual picture associated  with  a quantum critical point. In the conventional  scheme, \cite{Hertz76,Millis93,Moriya90,Moriya03}  nonFermi-liquid (NFL) features are expected to be confined to a fan-shape region  of the phase diagram, which  would here emerge   from the  zero temperature SDW critical point if superconductivity was absent.    Anomalies of the normal phase  are actually found on a much broader range of pressure and temperature, defining an   extended  quantum critical domain  that is intimately connected to the incursion  of Cooper pairing in the phase diagram.
  
In this work we address this  problem by  examining   the   single particle properties of the normal state   underlying the  SDW-SC   sequence of instabilities. The analysis is carried out  by the weak coupling renormalization group  method  (RG) in the framework of the   quasi-1D  electron gas  model,\cite{Nickel0506,Duprat01,Bourbon09} which is generic of  the close proximity of antiferromagnetism to superconductivity     in   low-dimensional compounds like the Bechgaard salts.

Previous one-loop RG calculations carried out  along these lines   have shown how the   interplay between antiferromagnetism and superconductivity relies on the interdependent pairings  of  electron-electron (Cooper) and electron-hole (density-wave or Peierls)  scattering channels, an interdependence that is present to every order in perturbation theory \cite{Duprat01,Nickel0506,Bourbon09}.  For repulsive intrachain interactions, the relative  stability of  SDW and d-wave SC (SCd) orders  can thus  be followed and the  essence of   instability lines defining  the phase boundaries   captured as a function of  nesting deviations of the Fermi surface (FS). These deviations are commonly parameterized by   the   next-to-nearest neighbors  transverse hopping, $t_\perp'$, which modulates the warping  of the  FS; its amplitude, thought much smaller than the nearest-neighbor interchain hopping term  $t_\perp (\sim 200$K), acts as a low energy scale that tunes nesting frustration and simulates   the primary effect  of  pressure\cite{Yamaji82,Montambaux88,Nickel0506,Duprat01}.  By raising $t_\perp'$, the SDW instability line, instead of touching zero temperature at a critical $t_\perp'^*$, terminates where the SCd line begins at a finite $T_c$. 
It follows that a conventional pressure driven SDW quantum critical  point is in a way  avoided by the presence of superconductivity   whose critical temperature  $T_c$  undergoes a steady decline as $t_\perp'$ increases away from $t_\perp'^*$\cite{Nickel0506,Duprat01}. 

The RG analysis has further shown that the impact of interfering pairings is not  limited to the sequence of   instability lines, but   stretches out   deeply in the metallic  state\cite{Bourbon09}. This takes on   particular importance    in  the superconducting sector of the calculated phase diagram. A sizable region of the normal phase  can thus  be defined  where SDW  correlations are reinforced by d-wave Cooper pairing,  under the guise of  a Curie-Weiss (CW) temperature variation of the SDW susceptibility above $T_c$. In contrast to the customary description  of quantum critical phenomena,\cite{Moriya03}  the  enhancement is found to persist  down to the lowest temperature with a strength that correlates with   the  size of the superconducting $T_c$.  It has been shown how this   mechanism of enhancement supplies  a coherent explanation of the   CW  temperature dependence of  $(T_1T)^{-1}$, \cite{Bourbon09} formerly observed   in   the low temperature metallic phase   of  different members of the Bechgaard salts series under pressure.\cite{Wu05,Shinagawa07,Brown08,CreuzetRMN,Bourbon84,Kimura11}

 In what follows, we  pursue   the above program a step further and study for the same  model conditions the single particle properties of the normal state. This is achieved by using the one-loop results for the scattering amplitudes in  the two-loop calculation   of the one-particle Matsubara  self-energy,   $\Sigma({\bf k}_F,i\omega_\nu)$,  on the FS. The calculation is carried out  by the RG method over the whole temperature interval    of the normal state,  that is  from the  high   temperature -- 1D -- domain  down to arbitrarily far below  the deconfinement temperature $T_X$  for normal electrons.  In   lowest order\cite{Bourbon84,Emery83},   $T_X \sim t_\perp$,  which  fixes the temperature scale under which  transverse  motion of electrons becomes quantum mechanically coherent and the electron system crosses  over to a two-dimensional behavior. In the case of  weakly dimerized chains  and relatively weak interactions  relevant to the Bechgaard salts, this corresponds to the broad, but rather unexplored  temperature region,\cite{Tsuchiizu07} where  anomalies of the normal state can occur and  
   instabilities of the metallic state against long-range order take place, these becoming ultimately affected by the presence of the much smaller scale $t_\perp'\ll t_\perp$.   
 
 The  momentum resolved quasi-particle weight $z({\bf k}_F)$ on the FS along with the renormalization factors for the perpendicular  hopping amplitudes  can thus be extracted   from the RG flow of  $\Sigma({\bf k}_F,i\omega_\nu)$. In  the   low temperature region where the normal phase  is subjected to instabilities of the electron gas,  the present analysis  demarcates  from previous works  based on mean-field\cite{Boies95,Essler02,Essler03} and infinite transverse dimensions treatments of $t_\perp$, \cite{Arrigoni99,Arrigoni00,Biermann01,Berthod06} in which the momentum dependent details of the FS warping are considered as virtually  irrelevant and the possibility of instabilities of the electron gas,  absent, in weak coupling.  In the high temperature region above $T_X$, namely  where the influence of $t_\perp$ is  perturbative, the results   show some concordance with these earlier investigations.

Our  analysis  is completed by performing from the Pad\'e approximants method, the  analytic continuation of  $\Sigma({\bf k}_F,i\omega_\nu)$   to its retarded form, $\Sigma^{\rm R}({\bf k}_F,\omega)$, defined on the real frequency axis $\omega$.    We   evaluate   one-particle  spectral quantities  on the FS. As a function  of energy and temperature,  we examine   deviations     from the Fermi liquid predictions  as    $t_\perp'$ tunes   the SDW-SCd sequence of  instabilities  in the phase diagram.    

A  particular emphasis is put on the momentum resolved electron-electron scattering rate, $\tau_{{\bf k}_F}^{-1}$,  on the FS, a key ingredient entering in  resistivity when Umklapp scattering is present. The superconducting region of the phase diagram discloses an anomalous behavior of $\tau_{{\bf k}_F}^{-1}$,  well captured by the superimposition of  a linear-$T$ and $T^2$ temperature dependences within  the CW domain of spin fluctuations.  The decay of the NFL  linear-$T$ component   alongside the emergence of a regular quadratic contribution  scale with the evolution of $T_c$ over a wide range of $t_\perp'$,  in accordance with an expanded  region of quantum criticality for spin fluctuations.  The  $\tau_{{\bf k}_F}^{-1}$ anisotropy around  the FS is a fingerprint of Cooper pairing that modifies the scattering of quasi-particles by  spin fluctuations  in  the normal state.

 In Sec.\,II, the quasi-1D electron gas model is introduced,   followed  in Sec.\,III by the  calculation  of one-particle Matsubara self-energy on the Fermi surface from the  one-loop scattering amplitudes. In Sec.\,IV, we proceed with the calculation of quasi-particle weight  and renormalization of the transverse hopping integrals. By means of analytical continuation, we examine different spectral properties of the normal state quasi-particles for the quasi-1D electron gas model. In Sec.~V, we discuss the results in connection with experiments in systems like the Bechgaard salts and conclude.

 \section{Model}

We consider the quasi-1D electron gas model for a linear array of $N_\perp$ conducting chains of length $L$. The partition function of the model can be  expressed as a functional integral
\begin{equation}
Z = \iint \mathfrak{D}\psi^*\mathfrak{D}\psi \ e^{S_0[\psi^*,\psi] + S_I[\psi^*,\psi]},
\end{equation}
 over the set of anticommuting fields $\{\psi_{p,\sigma}^{(*)}(\bar{k})\}$ for  right ($p=+)$ and left ($p=-)$ moving electrons  of spin $\sigma$ and momentum-frequency vector $\bar{k}=({\bf k},i\omega_\nu)$. Here   $\mathbf{k}=(k,k_\perp)$ and \hbox{$\omega_\nu =  (2\nu + 1)\pi T$} are  the wave vector and the fermion Matsubara frequency, respectively ($\hbar=1$, $k_B=1$, throughout). The free part, $S_0$, of the action is given by 
\begin{equation}
S_0[\psi^*,\psi] = \sum_{p,\bar{k},\sigma} [G_p^0(\bar{k})]^{-1}\psi^*_{p,\sigma}(\bar{k})\psi_{p,\sigma}(\bar{k}), 
\label{S0}
\end{equation}
where 
 \begin{equation}
G_p^0(\bar{k}) = [i\omega_\nu -E_p({\bf k})]^{-1},
\end{equation}
 is the one-electron bare  propagator. The    energy spectrum  of the model takes the form
\begin{equation}
\label{spectrum}
\begin{split}
E_p(\mathbf{k} ) =  v_F(pk-k_F)  -2t_{\perp }\cos k_\perp  & -2t'_{\perp}  \cos 2k_\perp,\end{split}
\end{equation}
where $v_F$ and $k_F=\pi/2$ are  the longitudinal Fermi velocity and wave vector;   $t_{\perp}$  is the  nearest-neighbor hopping integral  in the  perpendicular direction (the lattice constants are set to unity). The small transverse second nearest-neighbor hopping   $t_{\perp }'\ll t_{\perp }$ paramaterizes deviations to perfect  nesting  of the open  Fermi surface, which simulates the most important effect of the pressure in our model.  The  parameters of the  quasi-1D   spectrum are fixed in accordance to  the range  found in band calculations for the Bechgaard salts,   that is to say with a longitudinal Fermi energy \hbox{$E_F\simeq 15 t_{\perp } $  $\simeq 3000$K}, which is an order of magnitude larger than $t_\perp$ \cite{Grant82,Ducasse86,LePevelen01}; the bandwidth cutoff $E_0=2E_F$ ($\equiv \pi v_F)$,  is fixed  as twice the longitudinal Fermi energy. 

The interacting part, $S_I$, of the action takes the form
\begin{widetext}
\begin{equation}
\label{Hamiltonian}
\begin{split} S_I   =   -   {\pi v_F\over TLN_\perp} &\sum_{\{\bar{k},\sigma\}} \,  \big[\, g_{\{\sigma\}}(k'_{\perp1},k'_{\perp2};k_{\perp1},k_{\perp2}) \, \psi^*_{+,\sigma'_1}( \bar{k}'_1)\psi^*_{-,\sigma'_2}( \bar{k}'_2)\psi_{-,\sigma_2}( \bar{k}_2)\psi_{+,\sigma_1}( \bar{k}_1) \cr
   &\ \ \  \  \ + \ {1\over 2} g_{3}(k'_{\perp1},k'_{\perp2};k_{\perp1},k_{\perp2}) \, \psi^*_{+,\sigma_1}( \bar{k}'_1)\psi^*_{+,\sigma_2}( \bar{k}'_2)\psi_{-,\sigma_2}( \bar{k}_2)\psi_{-,\sigma_1}( \bar{k}_1)\cr   
& \ \ \  \  \ + \ {1\over 2} g_{3}(k'_{\perp1},k'_{\perp2};k_{\perp1},k_{\perp2}) \, \psi^*_{-,\sigma_1}( \bar{k}'_1)\psi^*_{-,\sigma_2}( \bar{k}'_2)\psi_{+,\sigma_2}( \bar{k}_2)\psi_{+,\sigma_1}( \bar{k}_1)\big]\delta_{\mathbf{k}_1+\mathbf{k}_2= \mathbf{k}'_1+\mathbf{k}'_2(\pm\mathbf{G}) },\end{split}
\end{equation}
\end{widetext}
which comprises a normal part 
\begin{equation}
\label{ }
g_{\{\sigma\}}(k'_{\perp1},k'_{\perp2};k_{\perp1},k_{\perp2}) = g_2\delta_{\sigma_2\sigma'_2}\delta_{\sigma_1\sigma'_1} -g_1\delta_{\sigma'_1\sigma_2}\delta_{\sigma'_2\sigma_1},
\end{equation}
  regrouping the bare backward ($g_1$) and forward ($g_2$) scattering amplitudes   between right and left moving electrons. The half-filling character of the band -- due to the small dimerization gap $\Delta_D$ of the chains -- gives rise to  Umklapp scattering \cite{Emery82}, of bare amplitude 
 \begin{equation}
 \label{Umklapp}
g_{3}(k'_{\perp1},k'_{\perp2};k_{\perp1},k_{\perp2}) \approx g_1\Delta_D/E_F, 
\end{equation}
 and  for which momentum conservation is satisfied  by adding    the longitudinal reciprocal lattice vector $\bar{G}= (\mathbf{G},0)$, where $\mathbf{G}= (4k_F,0)$.   All the   bare -- initial --  couplings are  normalized  by the bandwidth  $\pi v_F$. They are local or intrachain like and independent of the transverse momentum, the momentum dependence being generated by the RG flow.

Initial couplings can be seen as input phenomenological parameters of an effective theory of the electron gas. Following the example of Fermi liquid theory, their  values can be delimited for the most part by experiments. The temperature dependent enhancement of uniform magnetic susceptibility,   for example,  is known to be related to the backscattering amplitude \cite{Nelisse99,Fuseya07}; its observation is compatible with  the range  of values $g_1\approx 0.2-0.4$\cite{Wzietek93}.  The weak dimerization of organic stacks  observed by X-rays\cite{Thorup81} leads, according to band calculations, to  the small ratio $\Delta_D/E_F \sim 0.1$ for the dimerization gap,\cite{Grant83} which according to (\ref{Umklapp}) yields $g_3 \approx 0.02-0.04$. As for the range of $g_2$, it can be fixed by matching the   temperature scale of  the SDW instability  in       low pressure experiments, \cite{Jerome82,Klemme95} namely $T_{\rm SDW}\approx 10-25$K,   to  the predictions of RG calculations for $T_{\rm SDW}$ at small $t_\perp'$; one thus finds $g_2\approx 0.5-0.7$. Although   the above set of figures is merely indicative of what may occur in practice,  it turns out to be  rather generic of the succession of ground states that actually takes place in low dimensional  materials like the Bechgaard salts.

In the following,   we shall use  $g_1= g_2/2=0.32$ and $g_3=0.02$ as   initial couplings. There is nothing special about this choice other than it  coincides with the one  used in previous RG analysis for the phase diagram and NMR $T_1^{-1}$   \cite{Bourbon09,Nickel0506},  thus allowing a common   framework for the connection with experiments.

\section{Renormalization group  analysis}

\begin{figure}
 \includegraphics[width=5.0cm]{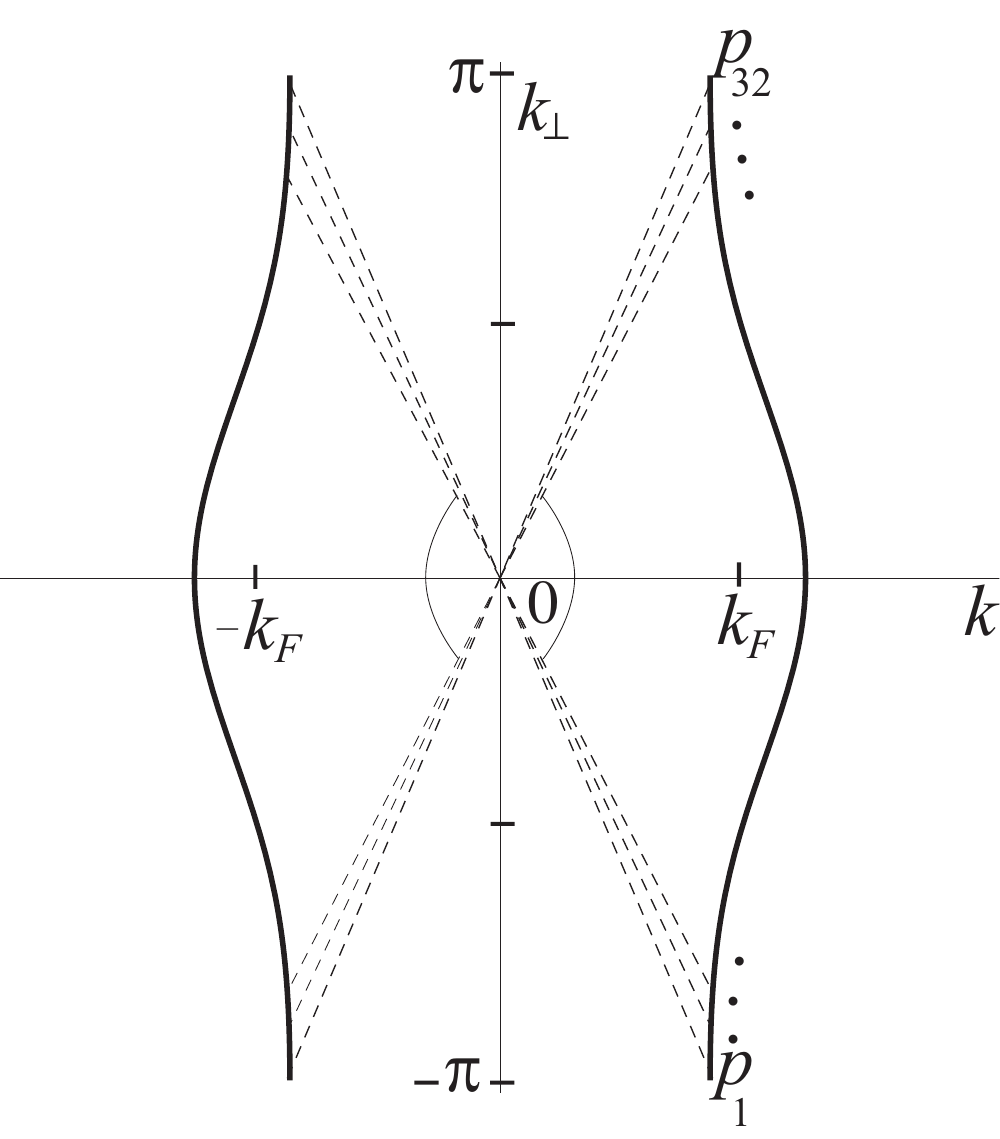}
 \caption{The Fermi surface of the quasi-one dimensional electron gas model. Each dashed line ends  to a patch $p_{i=1 ... 32}$  on each Fermi sheet. \label{FS}} 
 \end{figure}

We apply a Kadanoff-Wilson   RG approach   to the partition function $Z$\cite{Bourbon91,Bourbon03}.   In the present quasi-1D problem, this is achieved by     first dividing the  open Fermi surface (FS)     into 
32 pieces or patches  \cite{Zanchi97,Zanchi00},   each  centered on a particular transverse momentum $k_\perp$   that parametrizes  a     point  ${\bf k}_{F,p}=\big(k_{F,p}(k_\perp),k_\perp\big)$ on the FS  sheet,  $p=\pm$ (Fig.~\ref{FS}), where the longitudinal  component   $k_{F,p}(k_\perp)$ follows from  the  FS equation,  $E_p({\bf k}_{F,p})=0$. 

The RG  proceeds from successive  partial traces of $Z$ over a subset of variables $\bar{\psi}^{(*)}$ of the action in  an energy shell of thickness $\delta E_0(\ell)/2=E_0(\ell)d\ell/2$,  above and below  the FS. This is done in a perturbative scheme  in the energy intervals  \hbox{$\Delta_\pm=[\pm {1\over 2} E_0(\ell),\pm E_F]$},  ranging from the cutoff energy $E_F$ down to ${1\over 2} E_0(\ell) = {1\over 2}E_0 e^{-\ell}$,   above and below  the FS at the step $\ell\ge 0$. For a warped  Fermi surface like in Fig.~\ref{FS}, the RG integration   proceeds  with a cascade of  contractions of  three-particles  diagrams over the whole energy intervals $\Delta_\pm$. A momentum  dependence for the scattering amplitudes is generated, which is restricted to the set $\{k_\perp\}$ of three transverse  external variables  at  the energy distance $ {1\over 2}E_0(\ell)$ from the FS.    At the one-loop level for instance \cite{Sedeki10b}, the procedure is equivalent to the   one-particle irreducible  scheme of the functional RG\cite{Honerkamp00},    as  previously  applied to the model under study  in Ref.\cite{Nickel0506}. 

\subsection{Summary of one-loop results}
Before embarking on the study of one-particle properties, it is  useful to first single out  the main results of the previous RG calculations carried out  at the one-loop level,\cite{Bourbon09,Nickel0506} These refer to the particle and hole  excitations entering   in the   single particle  Matsubara self-energy at the two-loop level.

At the  one-loop level  the cascade of shell contractions,  ${\cal O}(S_I^2)$, in interactions yields the flow equations  shown diagrammatically in  Fig.~\ref{flow}-a  for the normal and Umklapp scattering amplitudes. The related  expressions were  obtained by Nickel {\it et al.}\cite{Nickel0506}  [Eqs.\:(10)-(12)   of Ref.~\cite{Nickel0506}], and  are summarized  as follows:  
 \begin{figure}
  \includegraphics[width=9.0cm]{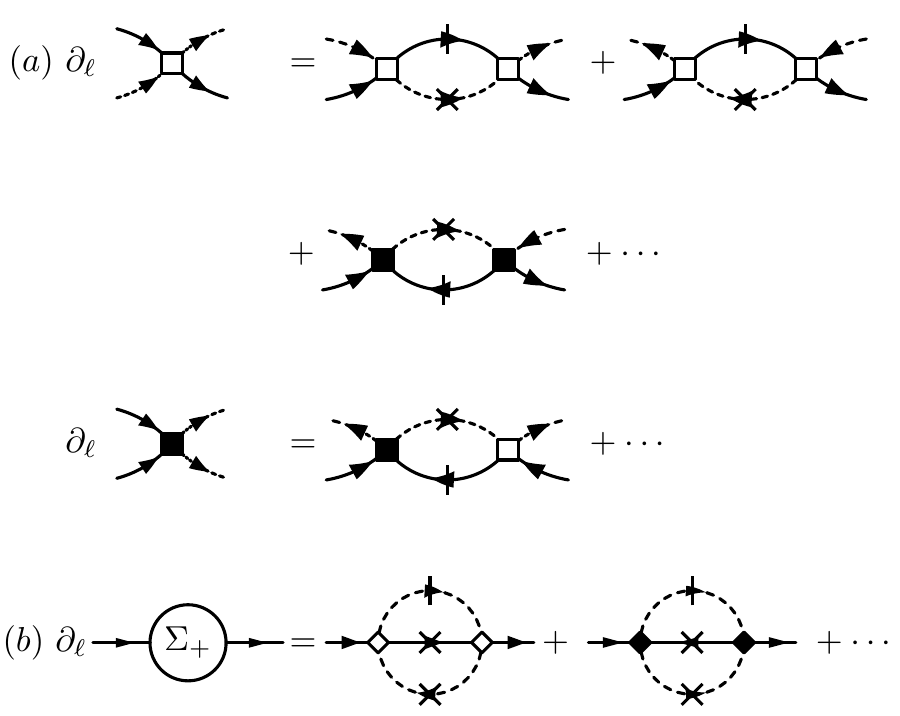}
  \caption{ Flow equations for :  (a) the $g_{1,2}$ (open square) and Umklapp $g_3$ (full square) scattering amplitudes for right (continuous line) and left (dashed line) moving electrons; (b) the one-particle Matsubara self-energy, $\Sigma_+$. The crossed and slashed  lines refer to the high energy intervals $\Delta_\pm$  and the last outer energy  shell, respectively (permutations between crossed and slashed line are not shown). \label{flow}} 
 \end{figure}
  \begin{eqnarray}
\label{gn}
 &&\!\!  \!\! \partial_\ell     {g}_{i=1,2}(k'_{\perp1},k'_{\perp2};k_{\perp1},k_{\perp2}) \!   =\!  \!\sum_{{\cal P}_{nn'}}\!\Big\{\epsilon^{n,n'}_{C,i}\! \!\left\langle {g}_n  \cdot {g}_{n'} \cdot \partial_\ell {\cal L}_C\right\rangle_{k_\perp}  \cr   
&&     
   \  \ \  \ \ \  \ \ \ \ \ \ \ \ \ \ \ \  \ \ \ \ \ \ \ \ \ \ \ \ \ \ \ +   \  \epsilon^{n,n'}_{P,i}\!\left\langle    {g}_n  \cdot {g}_{n'} \cdot \partial_\ell{\cal L}_P\right\rangle_{k_\perp} \! \Big\}, \cr\cr
& &\!\!  \!\! \partial_\ell {g}_3(k'_{\perp1},k'_{\perp2};k_{\perp1},k_{\perp2})      =    \sum_{{\cal P}_{3n}} \epsilon^{3,n}_{P,3} \,\left\langle \tilde{g}_3 \cdot \tilde{g}_{n}\cdot \partial_\ell{\cal L}_P\right\rangle_{k_\perp}, 
\end{eqnarray}
where the  $k_\perp$-dependence of  convolution products, constrained by momentum conservation,  has been  masked  for simplicity; here $\langle \ldots \rangle_{k_\perp}$ stands for an average over $k_\perp$. The Peierls and Cooper loops at finite $T$ read\cite{Nickel0506}
\begin{eqnarray}
&&\!\!\!\partial_\ell{\cal L}_{P,C}(k_\perp, q_{P,C})=  \sum_{\nu=\pm} \theta(|E_0(\ell)/2 + \nu {\cal A}_{P,C}|- E_0(\ell)/2 )\cr
&& \times {1\over 2}\left( \tanh{E_0(\ell)  +2 \nu {\cal A}_{P,C}\over 4T} + \tanh{ E_0(\ell)\over 4T}\right)\cr
&& \times {E_0(\ell)\over 4 E_0(\ell) + 2 \nu {\cal A}_{C,P}},
\end{eqnarray}
 which  are  evaluated at $q_P = k_{\perp1}-k_{\perp1}'$ and $q_C=k_{\perp 1}+k_{\perp 2}$, respectively; ${\cal A}_{P,C}(k_\perp,q_{P,C})= -E_+(k_F,k_\perp) \mp E_-(-k_F,k_\perp + q_{P,C})$ and   $\theta(x)$ is the step function.  For a given $g_i$, the related sum  in (\ref{gn}) collect  all  possibilities $\{{\cal P}_{nn'}\}$  of diagrams   belonging  to the  Peierls and Cooper scattering channels $P$  and $C$.     The coefficients $\epsilon_{C,P,i}^{nn'}$ fix  the   sign and spin multiplicity  of each loop contribution; it differs for closed loops    ($\epsilon_{P,i}^{nn'}=-2)$, vertex corrections ($\epsilon_{P,i}^{nn'}=1)$ and ladder  ($\epsilon_{C,i}^{nn'}=-\epsilon_{P,i}^{nn'}=-1$) diagrams.

The equation  for the normal part $g_{\sigma}(g_1,g_2)$ superimposes    Peierls and  Cooper  loops that will mix and interfere at every order following  the $\ell$-integration. As for the Umklapp amplitude, $g_3$, it only contains Peierls loops, but in accordance with (\ref{gn}),  connects in a key manner with Cooper pairing via a linear  coupling to  normal interactions (Fig.~\ref{flow}). 

 The integration of  RG  equations is conducted up to $\ell \to \infty$ yielding the  scattering amplitudes at temperature $T$, which stands  as a distinct parameter from   $E_0(\ell)$.  A singularity in the scattering amplitudes signals an instability of the metallic phase toward the formation of a collective state. Strictly speaking, for an effective two-dimensional electron gas, fluctuations would prevent the possibility of long-range order at finite $T$. Potential couplings  or/and single particle hopping in a third direction are therefore needed so that a true phase transition at finite temperature can occur. The one-loop transition temperature can be seen as a characteristic scale for the onset of short-range order in two dimensions, but enough representative of  the actual transition if three-dimensional effects were included.

As to  the precise  nature of the normal state  instability, it  is best provided from the most singular response function $\chi_\mu$, at the   `ordering' temperature $T_\mu$ of the vertex part (\ref{gn}).  In the repulsive sector  of  bare intrachain couplings,  only the SDW and SCd   susceptibilities display singularities  depending  on the strength of the antinesting parameter $t_\perp'$. \cite{Duprat01,Nickel0506,Bourbon09} 
  
  In the RG framework,  the expression for the normalized susceptibility, $\tilde{\chi}_\mu\equiv \pi v_F \chi_\mu$, is given by
\begin{equation}
\label{ }
\tilde{\chi}_\mu({\bf q}_{\mu,0})=2\int_\ell \langle f_\mu(k_\perp)z^2_\mu(k_\perp)\partial_\ell {\cal L}_\mu\rangle_{k_\perp} d\ell,
\end{equation}
whose $k_\perp$ average expression  depends on the  vertex renormalisation factors $z_\mu(k_\perp)$ and the form factors $f_{\rm SDW}=1$ and $f_{\rm SCd}=2\cos^2 k_\perp$ for the SDW and SCd channels, respectively \cite{Nickel0506,Duprat01}.  The $z_\mu'$s obey the flow equation
\begin{equation}
\label{zmu}
\partial_\ell z_\mu(k_\perp) = \langle (\partial_\ell{\cal L}_\mu ) {g}_\mu z_\mu({k}'_\perp)\rangle_{{k}'_\perp},
\end{equation}
  governed by the combinations of  momentum dependent couplings,   ${g}_{\rm SDW} ={g}_{2}(k_{\perp},{k}'_{\perp};k_{\perp}+ \pi,k_\perp'+\pi) + {g}_{3}(k'_{\perp},k_{\perp};k_{\perp}' + \pi,k_\perp+\pi)$  and ${g}_{\rm SCd} =-{g}_1(k'_{\perp},-k'_{\perp};-k_{\perp},k_\perp) -{g}_2(k'_{\perp},-k'_{\perp};k_\perp,-k_{\perp})$. A singularity in ${g}_{\rm SDW}\ ( {\rm resp.,\ }  {g}_{\rm SCd})$  gives rise to a singularity in $\tilde{\chi}_{\rm SDW}\  ({\rm resp.,\ } \tilde{\chi}_{\rm SCd})$ at the ordering temperature $T_{\rm SDW}$  ({\rm resp.,\ } $ T_c$).

\begin{figure}
 \includegraphics[width=8.0cm]{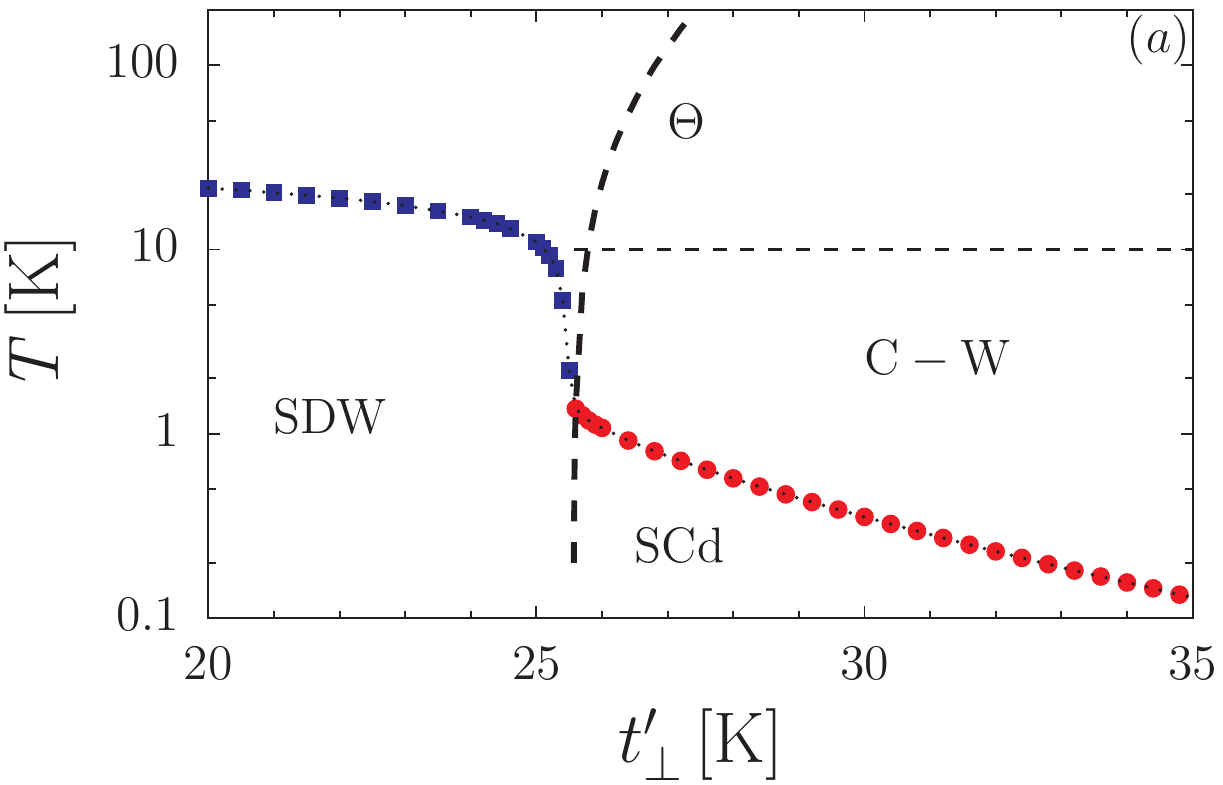} \\ \includegraphics[width=8.0cm]{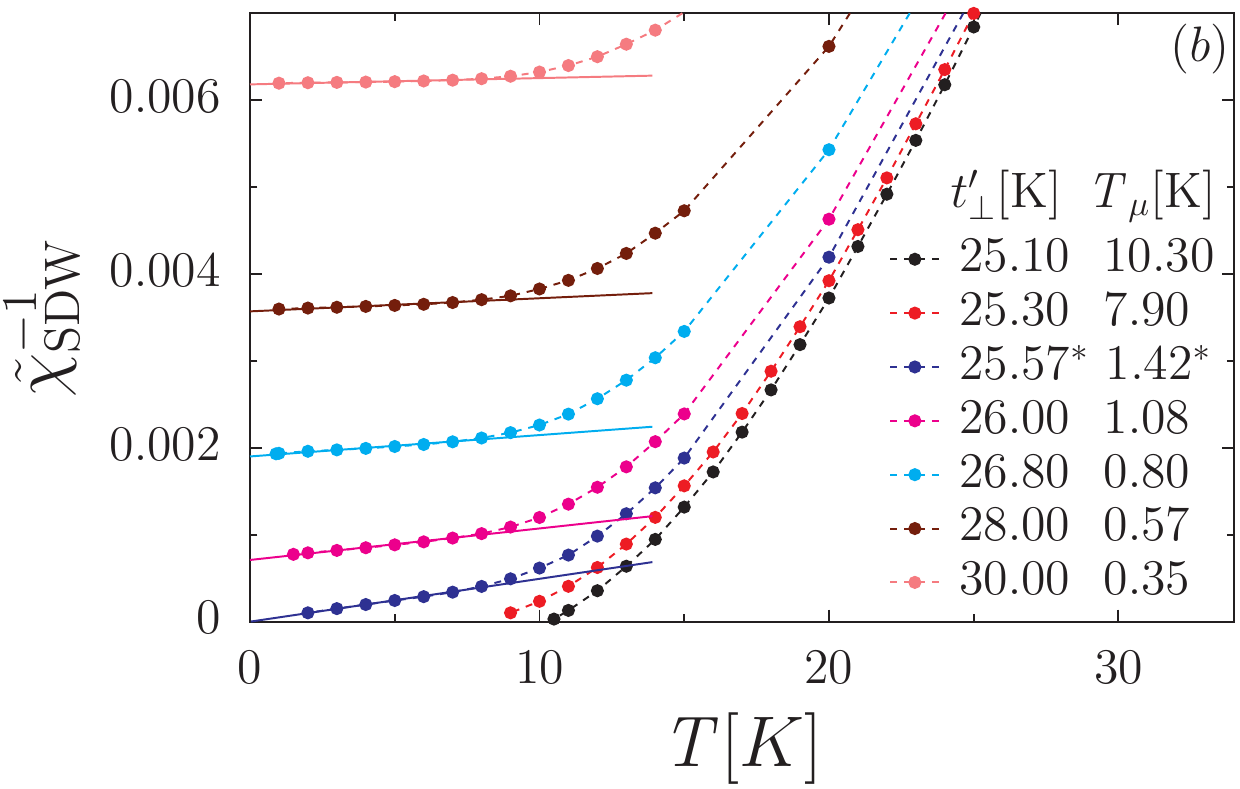}
 \caption{(Color on line) (a): Calculated phase diagram of the quasi-one-dimensional electron gas model from the renormalization group method  at the one-loop level; $T_{\rm CW}$ and  $\Theta$ define  the temperature region and   of the CW behavior for the inverse normalized SDW response function shown in (b) for different $t_\perp'$. \label{Phases}} 
 \end{figure}

Thus at small $t_\perp'$, nesting is sufficiently good so that a SDW instability occurs. It  is found with  a divergence  of the static SDW susceptibility, $\tilde{\chi}_{\rm SDW}({\bf q}_0)$,s  at the   temperature $T_{\rm SDW}$, when evaluated at the nesting vector  ${\bf q}_0=(2k_F, \pi)$ of $E_p({\bf k})$ in the absence of $t_\perp'$ (Fig.~\ref{Phases}). It is worth stressing here that this by no means indicates that the Cooper pairing channel is ineffective. On the contrary, the presence of electron-electron scattering and its interference with electron-hole  processes reduces considerably the temperature scale $T_{\rm SDW}$ for the magnetic instability.   For the parameters used here, for instance the $T_{\rm SDW}$  obtained  is reduced by a factor six or so compared to when the Peierls channel is singled out by putting to zero  the Cooper loops in (\ref{gn}).
\begin{figure}
 \psfrag{kp}{ $k_\perp$}
\psfrag{k2pp}{ $k'_\perp$}
\psfrag{Gs}{$g_1+g_2$}
\psfrag{G3}{$g_3$}
\includegraphics[scale=0.5]{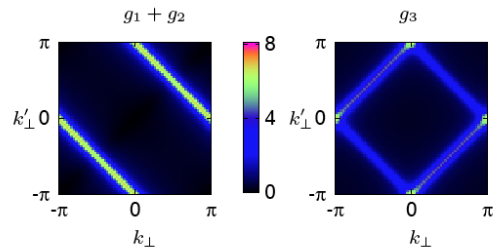} 
\caption{(Color on line) Momentum dependence of the projected  coupling constants $ {g}_1(k'_{\perp},-k'_{\perp};-k_{\perp},k_\perp) +{g}_2(k'_{\perp},-k'_{\perp};k_\perp,-k_{\perp})$ and $g_3(k'_{\perp},-k'_{\perp};k_{\perp},-k_{\perp})$   in the $(k_\perp,k_\perp')$ plane for the normal phase ($T= 25$K) of the SDW instability ($t_\perp'=25$K).\label{gSDW} }
\end{figure}

The signature of staggered density-wave correlations in the SDW sector of the phase diagram is also manifest in the momentum dependence of the coupling constants.  The Figure~\ref{gSDW}   shows typical momentum contour map of the normal combination  $ {g}_1(k'_{\perp},-k'_{\perp};-k_{\perp},k_\perp) +{g}_2(k'_{\perp},-k'_{\perp};k_\perp,-k_{\perp}) $, and Umklapp $g_3(k'_{\perp},-k'_{\perp};k_{\perp},-k_{\perp})$, as projected in the $(k_\perp,k_\perp')$ plane at zero incoming pair momentum. In the normal phase  ($T=25$~K) of the SDW instability at $t_\perp'=25$~K,  the amplitude of  both $g_1+ g_2$ and $g_3$    concentrates  essentially along the lines $k_\perp'=k_\perp \pm \pi$, namely at $\pi$ momentum transfer for staggered correlations along the transverse direction. Along the FS, as a function $k_\perp$, a  maximum of scattering  intensity for $g_1+ g_2$, though tiny,  is found  at the best nesting points $k_\perp=\pm \pi/4$  and $\pm 3\pi /4$, whereas for $g_3$, the maxima are at $k_\perp=0$  and    edges $ \pm \pi $  of the FS.

Now as $t_\perp'$ is tuned to larger values, nesting further deteriorates which gradually suppresses the Peierls  singularity of  the density-wave channel. As shown in Fig.~\ref{Phases}-a, $T_{\rm SDW} $ then first decreases monotonically until     the critical value   $t_\perp'^*$ ($\simeq 25.6$K for the parameters of the model) is approached where it sharply drops  (Fig.~\ref{Phases}). However,   in its fall, $T_{\rm SDW}$ does not  reach zero temperature, but ends at   the maximum $T_c$  of the SCd   instability line   ($ T_c^* \simeq 1.4$~K at $t_\perp'^*$),   along which the static d-wave susceptibility $\tilde{\chi}_{\rm SCd}(0)$ is singular at  zero Cooper  pair  momentum. Here  the influence of SDW correlations passes to the Cooper channel  by interference   and leads to d-wave pairing between electrons of opposite spins  on neighboring chains.\cite{Emery86,Caron86,Bourbon88}   

In the superconducting sector of the phase diagram where $t_\perp'\ge t_\perp'^*$, SDW fluctuations  display a characteristic change of regime  in  the normal phase.  This is apparent in the $T$ variation of $\tilde{\chi}_{\rm SDW}^{-1}$ of Fig.~\ref{Phases}-b, where the   temperature, $T_{\rm CW}\sim 10$~K, marks out the scale below which $\tilde{\chi}_{\rm SDW} $ is no longer  singular, but evolves toward a finite intercept as $T\to 0$. In the scaling language,    $T_{\rm CW}$  corresponds to the temperature range where the RG flow is no longer under the control of the SDW fixed point as a result of nesting deviations,  but crosses over to that of the SCd instability.\cite{Bourbon09}  Moreover, below $T_{\rm CW}$, $\tilde{\chi}_{\rm SDW}^{-1}$  evolves towards the $T\to 0$ intercept with a finite positive slope,   implying  that $\tilde{\chi}_{\rm SDW}({\bf q}_0)$ is still enhanced.  The SDW enhancement, though non singular,  is well captured by the Curie-Weiss  expression,
\begin{equation}
\tilde{\chi}_{\rm SDW}({\bf q}_0)\simeq {C \over T+ \Theta }, \ \ \ \ \ \ (T< T_{\rm CW}),
\end{equation}
 where $C,\Theta> 0$ (Fig.~\ref{Phases}-b). It follows that the SDW correlation length, $\xi\sim (T+\Theta)^{-1/2}$,   shows a regular reinforcement  down to the lowest temperature allowed by the calculations, namely $T_c$. 

Interestingly  enough, the CW enhancement turns out to be absent when the Cooper channel  is switched off and  all the Cooper loops in (\ref{gn}) or Fig.~\ref{flow}-a are put equal to zero. In the absence of Cooper pairing, the slope of $\tilde{\chi}_{\rm SDW}^{-1}$ is vanishingly small and $\tilde{\chi}_{\rm SDW}^{-1}$ is essentially constant below $T_{\rm CW}$ at all $t_\perp'\ge t_\perp'^*$. According to Fig.~\ref{Phases}-b when Cooper pairing is included, it is only when $t_\perp'$ is sufficiently far from $t_\perp'^*$ or $T_c$   small enough compared to $T_c^*$ that the slope $1/C$ tends to that of   the single channel limit.  The origin of CW SDW enhancement is a  direct consequence of the positive feedback of the Cooper (d-wave ) pairing channel on the Peierls one through the linear coupling to   Umklapp scattering [Eqs.~(\ref{gn})].  
\begin{figure}
 \psfrag{kp}{ $k_\perp$}
\psfrag{k2pp}{ $k'_\perp$}
\psfrag{Gs}{$g_1+ g_2$}
\psfrag{G3}{$g_3$}
\includegraphics[scale=0.5]{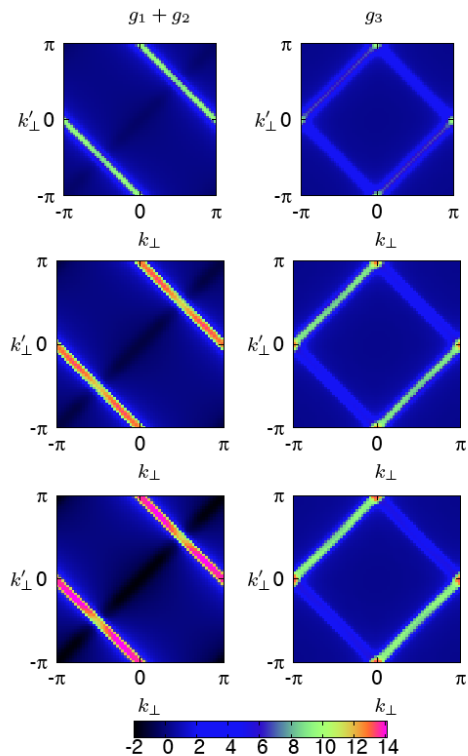}
\caption{(Color on line) Low temperature momentum dependence of the    coupling constants $ {g}_1(k'_{\perp},-k'_{\perp};-k_{\perp},k_\perp) +{g}_2(k'_{\perp},-k'_{\perp};k_\perp,-k_{\perp})$ (left) and $g_3(k'_{\perp},-k'_{\perp};k_{\perp},-k_{\perp})$ (right)    in the $(k_\perp,k_\perp')$ plane for the normal phase of the SCd instability ($t_\perp'=29$K);  $T=15$~K (top), 8~K (middle) and 4~K (bottom).\label{gSCd} } 
\end{figure}

Reinforcement of SDW correlations by Cooper pairing in the CW domain can  also be seen in the temperature dependent intensity  of scattering amplitudes in momentum space. In Fig.~\ref{gSCd},  the low temperature evolution of  $g_1 + g_2$  and $g_3$ is shown above and below  $T_{\rm CW}$ in the SCd region of the phase diagram at $t_\perp'=29$~K ($T_c=0.4$~K). Above $T_{\rm CW}$ (top of Fig.~\ref{gSCd}), the momentum profiles of $g_1+ g_2$ and $g_3$ at zero incoming pair momentum display  structures similar to those of the normal phase of the SDW instability (Fig.~\ref{gSDW}). However,   following the entry in the CW domain, the scattering intensity of $g_1+ g_2$ at   momentum transfer $q_\perp=\pm \pi$ increases continuously as the temperature is lowered over  the whole antinodal region of the FS at $k_\perp\ne \pm \pi /2 $. The increase  is indicative of a regular increase of staggered density-wave correlations which is  concomitant to a growing attractive component along the diagonal $k_\perp\sim k_\perp'$. This  is distinctive of constructive interference  between the Peierls and Cooper scattering channels. As shown in Fig.~\ref{gSCd}, the interference leads to an increase of Umklapp scattering at large momentum transfer, in particular for $k_\perp  \sim 0$ and $k_\perp \sim \pm \pi$, which contributes to the regular increase of SDW correlations in the CW temperature domain (Fig.~\ref{Phases}-b).

Regarding the CW scale  $\Theta$, it  rises from zero at $t_\perp'^*$  and increases rapidly as    $t_\perp'$ gets larger or $T_c$ decreases (Fig.~\ref{Phases}-a). The rapid growth of  $\Theta$ reflects  the  decrease of  spin fluctuations amplitude as a function of $t_\perp'$, as shown by the decline of zero temperature intercept of $\tilde{\chi}_{\rm SDW}({\bf q}_0,T\to 0)$ ($= C/\Theta$) in Fig.~\ref{Phases}-b.     It can be identified with  a characteristic energy scale for SDW fluctuations that softens and  becomes   critical as  $t_\perp'^*$   is approached    from above in the phase diagram.  As a function of  the tuning parameter $t_\perp'$, $\Theta$ evolves indeed according the critical form
\begin{equation}
\Theta \approx A(t_\perp'-t_\perp'^*)^{\eta}, 
\end{equation}
which  follows a linear variation, namely $\eta\simeq 1$, over the whole interval  $t_\perp'>t_\perp'^*$.  Attempting a quantum scaling approach to $\eta$, one can write $\eta = \nu z$, so that linearity is compatible   with the one-loop exponents    $\nu=1/2$ for   correlation length and $z=2$  for   dynamics of SDW fluctuations.  It has been checked that within one-loop approximation, linearity is essentially independent of the  initial  values of coupling constants.  The slope  $A$, however, is not universal: a  small increase in the initial value of Umklapp term, $g_3$, for instance, reduces appreciably  the amplitude of $A$ and then the rise     of $\Theta$  as a function of $t_\perp'$; this increases the CW slope $1/C$ for $\tilde{\chi}_{\rm SDW}^{-1}$  and  the amplitude of   spin fluctuations above $T_c$. 

\subsection{Two-loop one-particle self-energy}
\label{Self}
The successive  partial traces of $Z$ in the RG transformation yield a cascade of contractions  for the   $n\ge 3$ particles  vertex functions that are generated at each step of procedure.  For the 3-particles vertex, two  cascades of contractions for two internal fermion lines, followed by a  third line contraction on the final shell,  yield at every step  the two-loop corrections of one-particle  self-energy,  $\delta \Sigma_p$ \cite{Bourbon03,Chitov03,Sedeki10b}.  Details of the  $\delta \Sigma_p$ calculation     are given in Appendix~A. 

In the following, we shall  focus on the corrections evaluated  on the (bare) FS where the inverse one-particle propagator of  $S_0$ at the  step $\ell$ of  the RG, is of the form 
 \begin{equation}
 [G_p^0( {\bf k}_{F,p},\bar{k}_\perp)]_\ell^{-1} =   [z(\bar{k}_\perp)G_p^0({\bf k}_{F,p},\bar{k}_\perp)]^{-1} + z_\perp(\bar{k}_\perp),
\label{G}
\end{equation}
 where   $ \bar{k}_\perp=(k_\perp,i\omega_\nu)$. In accordance with the results of Appendix~A, the  one-particle self-energy corrections   give rise to  two renormalization factors,   $z(\bar{k}_\perp)$ and $z_\perp(\bar{k}_\perp)$, submitted to the initial conditions $z(\bar{k}_\perp)=1$ and $z_\perp(\bar{k}_\perp)=0$ at $\ell=0$, for all $\bar{k}_\perp$.

  Both $z$ and $z_\perp$ are real quantities, which from Eq.~(\ref{DG}) of the Appendix~A,     obey the following flow equations
\begin{equation}
  \begin{split}
  \label{zeq}
  &\partial_\ell \ln z(\bar{k}_{\perp}) =   {1\over 2}\iint \frac{d\,k'_\perp}{2\pi}\,\frac{d\,q_\perp}{2\pi}\cr
& \times  \Big\{ \Big(\! g_2({\widetilde k}_{\perp1})g_1({\widetilde k}_{\perp1}) -g^2_2({\widetilde k}_{\perp1})-g^2_1({\widetilde k}_{\perp1}) \Big)\, {I_1({\widetilde k}_{\perp},i\omega_\nu)}\cr
  &-\Big(g^2_2({\widetilde k}_{\perp2})+g^2_1({\widetilde k}_{\perp2})-g_2({\widetilde k}_{\perp2})
  g_1({\widetilde k}_{\perp2})\Big)\, {I_2({\widetilde k}_{\perp},i\omega_\nu)}\cr
  &-\Big(g^2_3({\widetilde k}_{\perp3})+g^2_3({\widetilde k}_{\perp4})
  -g_3({\widetilde k}_{\perp3})g_3({\widetilde k}_{\perp4})\Big)\, {I_3({\widetilde k}_{\perp},i\omega_\nu)}\Big\},
\end{split}
\end{equation}
and
\begin{equation}
  \begin{split}
  \label{zpeq}
 &  \partial_\ell z_\perp( \bar{k}_{\perp})=[z(\bar{k}_{\perp})]^{-1}
   {1\over 2}\iint \frac{d\,k'_\perp}{2\pi}\,\frac{d\,q_\perp}{2\pi}\cr
 & \times \Big\{ \Big(\! g_2({\widetilde k}_{\perp1})g_1({\widetilde k}_{\perp1}) 
   -g^2_2({\widetilde k}_{\perp1})-g^2_1({\widetilde k}_{\perp1}) \Big)\,  I'_1({\widetilde k}_{\perp},i\omega_\nu) \cr
   &-\Big(g^2_2({\widetilde k}_{\perp2})+g^2_1({\widetilde k}_{\perp2})-g_2( {\widetilde k}_{\perp})
   g_1( {\widetilde k}_{\perp2})\Big)\, { I'_2( {\widetilde k}_{\perp},i\omega_\nu)}\cr
   &-\Big(g_3^2({\widetilde k}_{\perp3})+g^2_3({\widetilde k}_{\perp4})
   -g_3( {\widetilde k}_{\perp3})g_3({\widetilde k}_{\perp4})\Big)\, {I'_3( {\widetilde k}_{\perp},i\omega_\nu)}\Big\},
  \end{split}
\end{equation}
where       ${\widetilde k}_{\perp i}$ and ${\widetilde k}_{\perp}$ are defined in (\ref{ktildei}) and (\ref{ktil}), respectively.
The  temperature dependent   coefficients $I_i$ and $I_i'$ are given by (\ref{I}-\ref{Ip}) and (\ref{I23}) of   Appendix A. The integration of (\ref{zeq}-\ref{zpeq}), together  with (\ref{gn}), is carried out up to $\ell \to\infty$, which yields the $z$ and $z_\perp$  factors at  temperature $T$. It worth stressing that both two-loop and one-loop diagrams are calculated using   free propagators (see Appendix~A). 

\section{Results}

\subsection{Quasi-particle weight} 
\label{zSec}

 \begin{figure}
 \includegraphics[width=9.0cm]{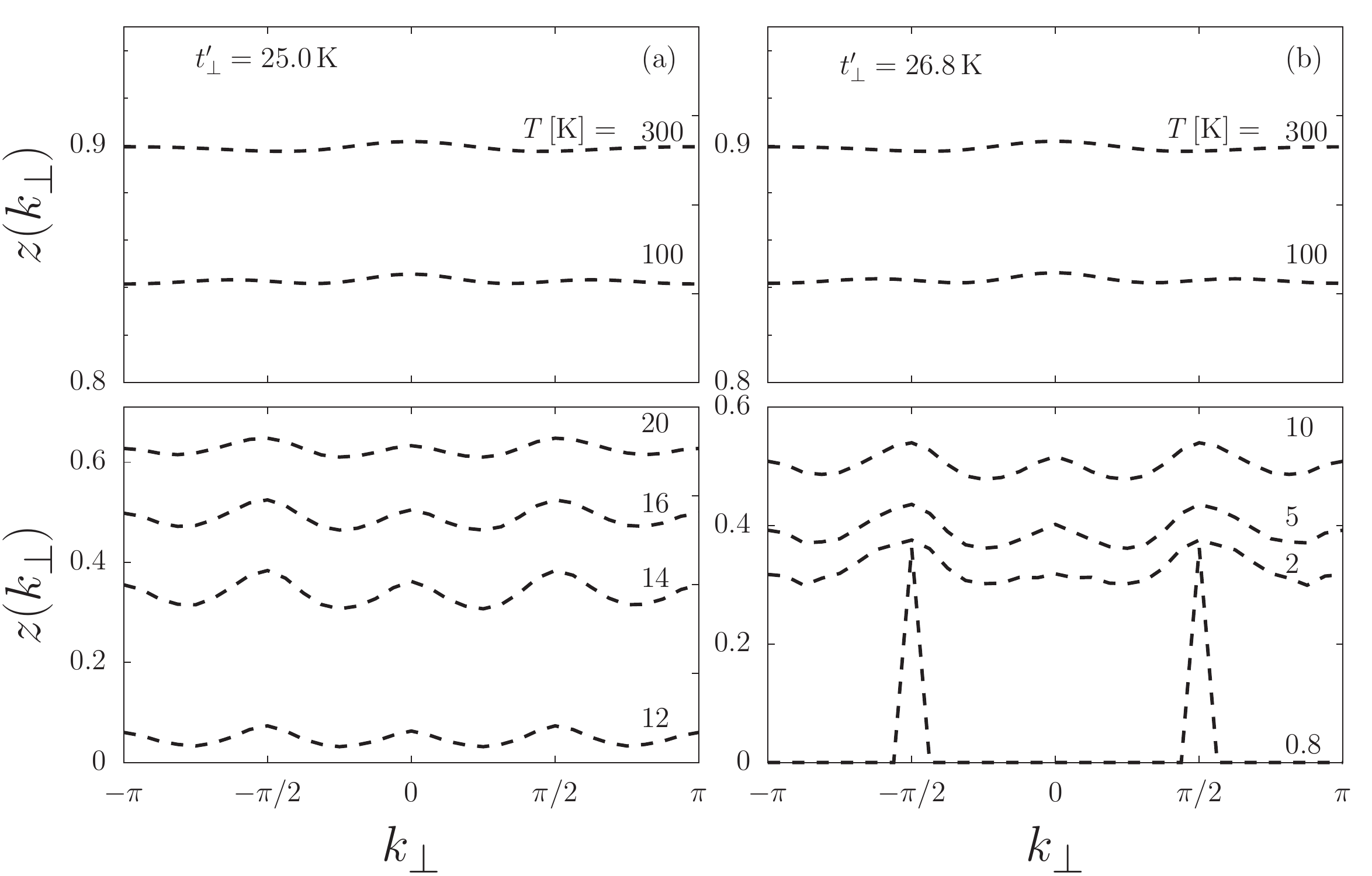}
 \caption{Variation of the quasi-particle weight on the Fermi surface as a function of $k_\perp$ at different temperatures. (a): SDW  ($t_{\perp}'= 25~{\rm K}< t_\perp'^*, T_{\rm SDW}\simeq 12~{\rm K}$); (b):  SCd ($t_{\perp}'= 26.8~{\rm K}> t_\perp'^*, T_c = 0.8~{\rm K}$). \label{zk} } 
 \end{figure}
An important quantity entering in the description of quasi-particles and that can be  extracted from the one-particle self-energy is the `momentum resolved' quasi-particle weight,  $z(k_\perp)\equiv z\big({\bf k}_{F,p}(k_\perp),i\omega_{\nu =0}\big)$,   on the  FS. It is obtained from the solution of Eq.~\ref{zeq} at $\omega_{\nu=0}=\pi T$. Typical $k_\perp$ and $T$ dependences  of $z(k_\perp)$     in the SDW and SC parts of the phase diagram  are shown in Fig.~\ref{zk}. 

\begin{figure}
 \includegraphics[width=7.0cm]{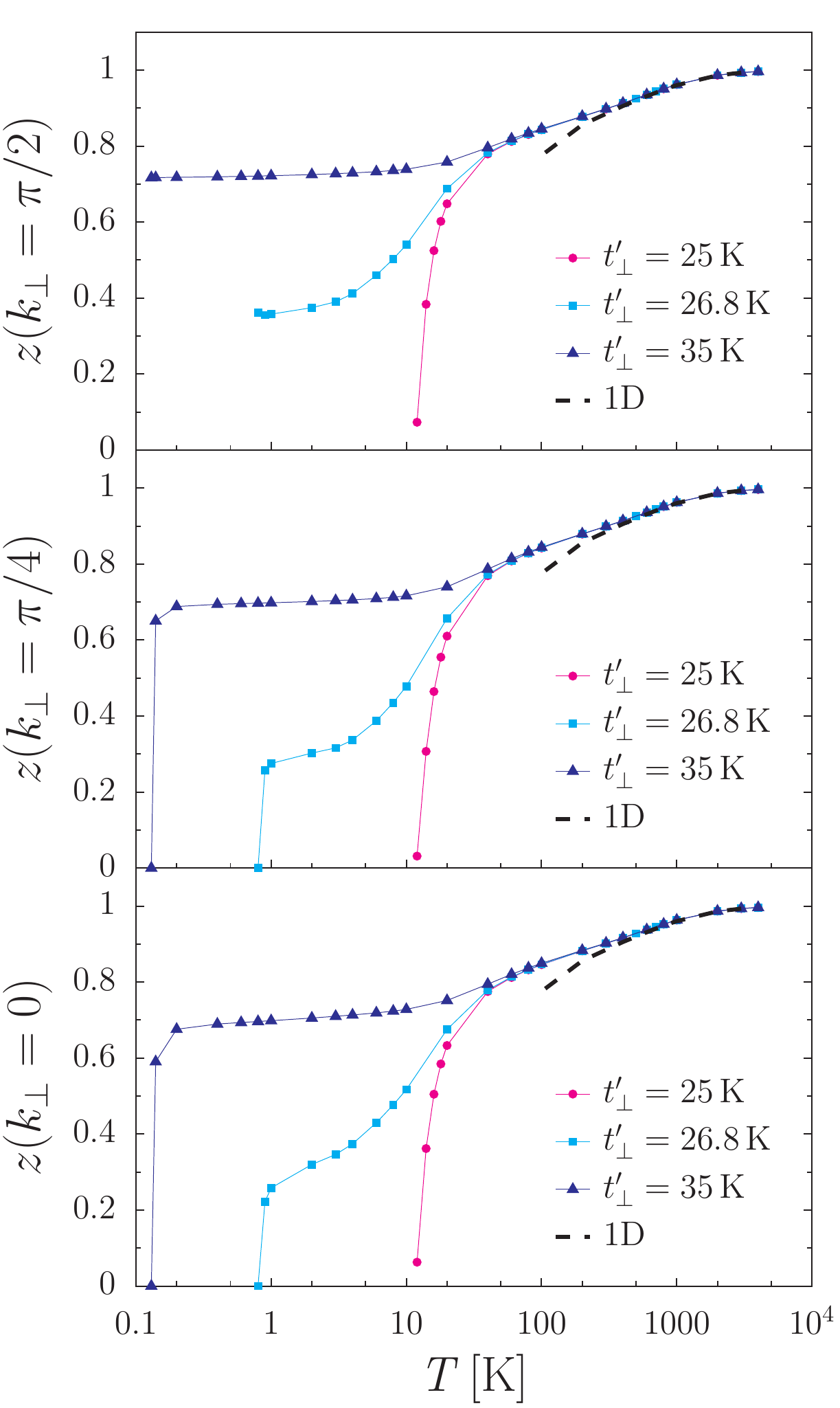}
 \caption{ (Color on line) Quasi-particle weight $z(k_\perp)$  as a function of temperature (log. scale)  and  antinesting parameter $t_\perp'$ for  $k_\perp=0$, $\pi/4$, and $\pi/2$. The dashed line corresponds to the power law  decay $z_{\rm 1D}$  of the  one-dimensional limit.\label{z} } 
 \end{figure}

In the SDW region and for the 1D temperature domain, namely  well above the scale of one-particle transverse coherence, $T_X\sim t_\perp$, $z(k_\perp)$ is only weakly reduced and displays little minima at  $k_\perp\sim \pm \pi/2$. According to (\ref{spectrum}), the $t_\perp$ part of the spectrum vanishes at those points. From a perturbation viewpoint of $t_\perp$, this  implies that 1D  effects are the strongest there. Such a  high temperature  modulation, thought small here, agrees with the results of earlier investigations based on perturbative and  mean field treatments of $t_\perp$ \cite{Boies95,Essler02,Essler03,Arrigoni99,Arrigoni00,Biermann01,Berthod06}, which  find the same location for the   spectral weight minima   on the FS.   Fig.~\ref{z} further  shows that in this 1D domain, $z(k_\perp)\sim T^{\alpha}$ and   decays as a non universal power law in temperature \big($\alpha\sim{\cal O}(g^2) \big)$,  in  accordance with the  summation of next-to-leading   logarithmic singularities  for the  self-energy, a well known result  of the   1D electron gas model.\cite{Solyom79,Bourbon91,Giamarchi04}(dashed line of Fig.~\ref{z}).

 As  $T$ goes below $T_X$, the influence of the $k_\perp$-dependent warping of the FS  becomes clearly non perturbative. The temperature decay of $z$, thought still present, becomes  less rapid than a  power law as previously noticed in previous RG investigations. \cite{Honerkamp03,Tsuchiizu07} This results from  the weakening of the two-loop singularity  and  modification of   Cooper and Peierls channels  interference  by  coherent warping of the FS. However,   as    discussed  in more details below,  when temperature  is further decreased toward the instabilities lines, but still in the normal phase, the quasi-particle weight decay develops quite distinctive features as a function of the antinesting parameter $t_\perp'$. The influence of the latter   becomes in its turn non perturbative, especially in the normal phase of the superconducting domain of the phase diagram.

 Below $T_X$, the position of  minima  
in $z(k_\perp)$ gradually shifts to $k_\perp=\pm \pi/4$ and $k_\perp=\pm 3\pi/4$, as the details of the Fermi surface become coherent. The shift  can be understood from   the  nesting condition of the whole spectrum, which reads
   \begin{equation}
\label{ }
E_+({\bf k}+{\bf q}_0) = -E_-({\bf k}) + \delta(k_\perp).
\end{equation}
Here \hbox{$\delta(k_\perp)= 4t_\perp'\cos 2k_\perp$} is the $k_\perp$-dependent deviation  from perfect nesting. The minima coincide with the loci, \hbox{$k_\perp=\pm \pi/4$}, and $\pm 3\pi/4$ on the FS where $\delta(\pm \pi/4)$ and $\delta(\pm  3\pi/4)$ vanish and  perfect nesting conditions prevail. Conversely, at $k_\perp =\pm\pi/2$, $0$, and $\pm \pi$,    $\delta(k_\perp)=\pm  4 t_\perp'$, and nesting deviations, like $z(k_\perp)$, are  maximum below $T_X$. 
As the temperature comes close to $T_{\rm SDW}$, $z(k_\perp)$ enters in the critical SDW regime and falls off toward zero  due  to singular scattering amplitudes  which signal the breakdown of the perturbative RG procedure (Figs.~\ref{zk} and \ref{z}).

Considering now the SC domain of the phase diagram at $t_\perp'> t_\perp'^*$, the  variation of $z(k_\perp)$ with respect to both $k_\perp$ and $T$ is similar to the one encountered in the SDW case   at high $T\gg t_\perp'$. As $T$  decreases well below $t_\perp'$, however,  $z(k_\perp)$ behaves differently.

Let us first examine the  $k_\perp$ dependence.  From Fig.~\ref{zk}-b, the modulation of $z(k_\perp)$ in the normal state is roughly similar to that obtained in the SDW case  
 with a slight modulation in $k_\perp$ that persists down to the CW regime (Fig.~\ref{zk}). In this  temperature  domain and down to $T_c$, the system is governed by the SCd fixed point and  the maxima at $k_\perp=\pm \pi/2$ remain stationary   at the approach of $T_c$, consistently with the location of  nodes for the SCd gap below $T_c$. The amplitude of the peaks at $k_\perp=0$ and $\pm \pi$ present a steady  reduction as the temperature is lowered. 

Turning  to  the $T$ dependence in the SCd sector  of the phase digram. We see from Fig.~\ref{z} that not too far from  $t_\perp'^*$ at  the intermediate  $t_\perp'=26.8$~K,      $z(k_\perp)$ exhibits a significative decay in the   metallic state extending far above the critical domain,  comprising the entire CW regime of   spin correlations. 
The amplitude of  the $z$ decay     correlates with the one  of SDW fluctuations in the normal phase as $t_\perp'$ is tuned away from $ t_\perp'^*$.  When  $t_\perp'$ is  far upward, at 35~K, for instance, one has $T_c=0.12$~K $\ll T_c^*$,  and a very weak temperature dependence is found for $z$ above $T_c$, which is very close to that of a Fermi liquid   (Fig.~\ref{z}).

If we now concentrate on the CW  temperature interval, $T_c\lesssim T \le T_{\rm CW}$,  we see that the $z$ decay at $k_\perp=0$ for instance,  can be seen as a result of  logarithmic corrections   that conform to the expression,  
\begin{equation}
\label{mfl}
z(k_\perp) = {z_0(k_\perp)\over 1 + g(k_\perp) \ln( T_{\rm CW}/T)},
\end{equation}
which is  compatible with the  phenomenology proposed by the marginal Fermi liquid  theory for the quasi-particle self-energy in two dimensions.\cite{Varma89} Here   $z_0(k_\perp)= z(k_\perp,T_{\rm CW})$ is the quasi-particle weight at the cutoff temperature $T_{\rm CW}=10$K, and $g(k_\perp)$ is the square of an effective  (averaged over two momentums variables) coupling constant normalized by $\pi v_F$.
 \begin{figure}
 \includegraphics[width=7.0cm]{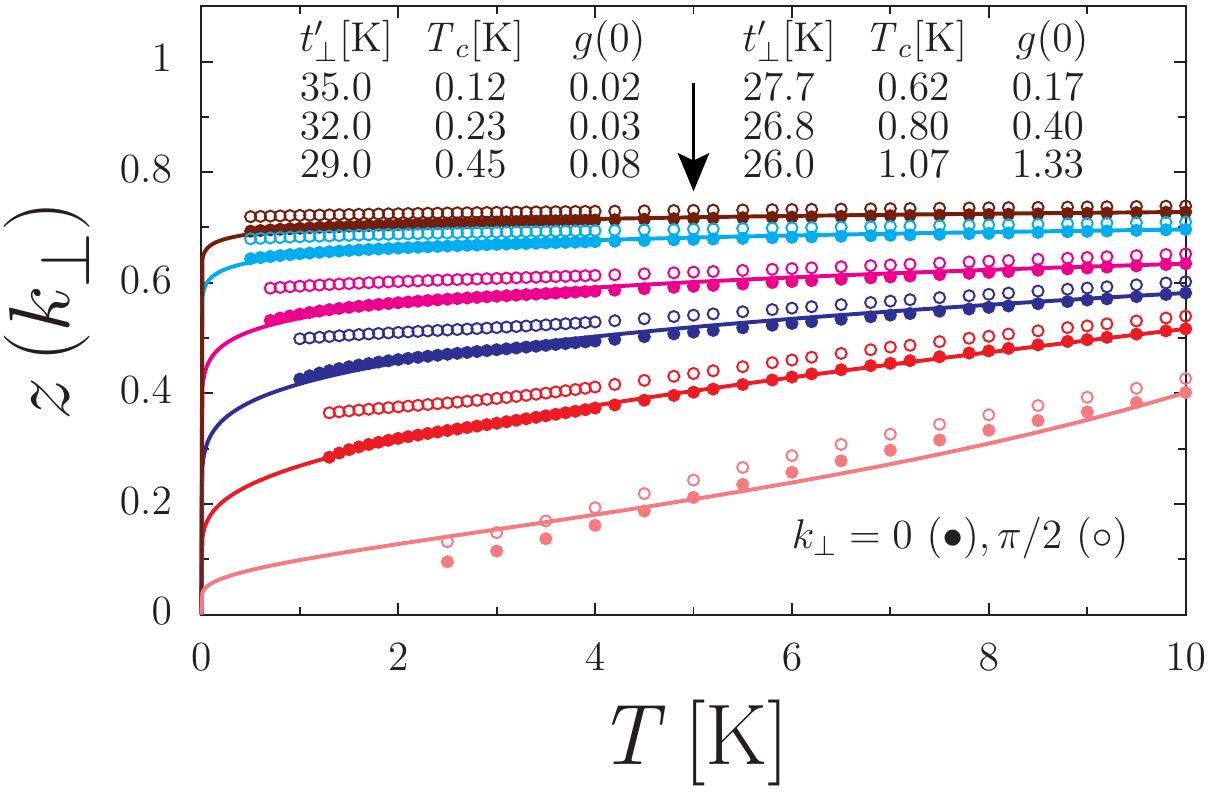}
 \caption{(Color on line) Low temperature dependence quasi-particle weight $z(k_\perp)$ for different antinesting parameters $t_\perp'$ for  $k_\perp=0$ (full circles),  and nodal point $\pi/2$ (open circles). The continuous lines correspond  to a fit to    Eq.~(\ref{mfl}) of the marginal liquid  theory  at $k_\perp=0$.\label{zmfl} } 
 \end{figure}

For most values of $t_\perp'$ shown  in Fig.~\ref{zmfl}, the  RG results in the CW regime  for $z$ at $k_\perp=0$ are compatible with an effective marginal Fermi liquid behavior, characterized by a weak coupling value for $g(k_\perp=0)$. As we will see in more details below (Sec.~\ref{Rate}), such a behavior is   in line with the presence of a $T$-linear  component in the temperature dependent electron-electron scattering rate for the same values of $t_\perp'$; a similar connection between $z$ and the scattering rate is well known in the context of   marginal Fermi liquid theory.\cite{Varma89}   According to Fig.~\ref{zmfl}, by raising $t_\perp'$  the reduction of $z$  declines, but the negative curvature is apparently maintained.  In the limit of  large $t_\perp'$ or  small $T_c$,   $z$  becomes ultimately constant in temperature with $g(0) \to 0$, as expected for a Fermi liquid.   

From Fig.~\ref{zmfl}, however, the quality of the fit  wanes significantly near the  critical value $t_\perp'^*$. At $t_\perp'=26$K for example, the reduction of $ z$  becomes rather important at low temperature with a  $g(0)$ value exceeding unity, as a result of large positive feedback of Cooper pairing on SDW fluctuations that lead the scattering amplitudes to flow to very strong coupling.  This signals that using one-loop RG calculations  in the one-particle two-loop self-energy may become  insufficient.  We will see indeed in Sec.~\ref{Rate} that the  pronounced reduction in $z$  coincides with a marked rising of the scattering rate  showing a progressive suppression of quasi-particles  akin to a pseudogap behavior.\cite{Bergeron11}  A full two-loop RG calculation of both self-energy and scattering amplitudes is required to overcome this difficulty at $t_\perp'\approx t_\perp'^*$.\cite{Sedeki10b}

Not too close to $t_\perp^*$, the marginal Fermi liquid fit works equally well for all $k_\perp$, with the exception of  the nodal points, $\pm \pi/2$, of the superconducting SCd gap below $T_c$. As shown in Fig.~\ref{zmfl}, there is a small positive curvature that  develops for $z(\pi/2)$ as $T$ approaches $ T_c$. A  Fermi liquid  component  is forming at those nodal points where  $z(\pm \pi/2)\to\, $constant    (Fig.~\ref{zmfl} and Figs.~\ref{z},\ref{zk}-b).   

 \subsection{Renormalization of transverse hopping integrals }
The  corrections produced by the flow of self-energy   generate the new term, $z_\perp(\bar{k}_\perp)$, in the single particle propagator (\ref{G}) (see also App.~A).  In the static limit, $z_\perp(k_\perp)\equiv z_\perp(k_\perp,\omega_{\nu=0})$ affects    the transverse part of the electron spectrum and then deforms the Fermi surface. According to (\ref{G}), the modification of the transverse energy is of the form,  $\delta \varepsilon_\perp(k_\perp)= z(k_\perp)z_\perp(k_\perp)$, which as an even function of $k_\perp$ can be     decomposed into the  Fourier series
\begin{equation}
\label{eperp}
\begin{split}
\delta \varepsilon_\perp(k_\perp) = \ t_0 & + \delta t_\perp \cos k_\perp +  \delta t_\perp' \cos 2k_\perp \cr
 & + \delta t_\perp'' \cos 3k_\perp + \ldots\ .
 \end{split}
\end{equation} 
The series coefficients consist of corrections to the chemical potential ($t_0$),   the first  ($\delta t_\perp$), second ($\delta t_\perp'$), third ($\delta t_\perp''$) $\ldots$,  nearest-neighbors  hopping integrals of the transverse direction. Here  $\delta t_\perp''$, as well as higher order harmonics are new terms of the spectrum generated by the RG flow. It turns out, however, that in all cases,  the total correction   $\delta \varepsilon_\perp$  is  very accurately  fitted to the existing form  of the transverse part of the spectrum (\ref{spectrum}), which involves  $\delta t_\perp$ and $\delta t_\perp'$ alone; the values of  $t_0$, $\delta t_\perp''$, etc .., are vanishingly small and can thus be disregarded as irrelevant terms. 

Adding both corrections to the spectrum    leads to the  temperature dependent renormalisation of  interchain hoppings, which become $\bar{t}_\perp= t_\perp + \delta t_\perp $  and  $\bar{t}_\perp'= t_\perp' + \delta t_\perp'$. 
 \begin{figure}
 \includegraphics[width=8.5cm]{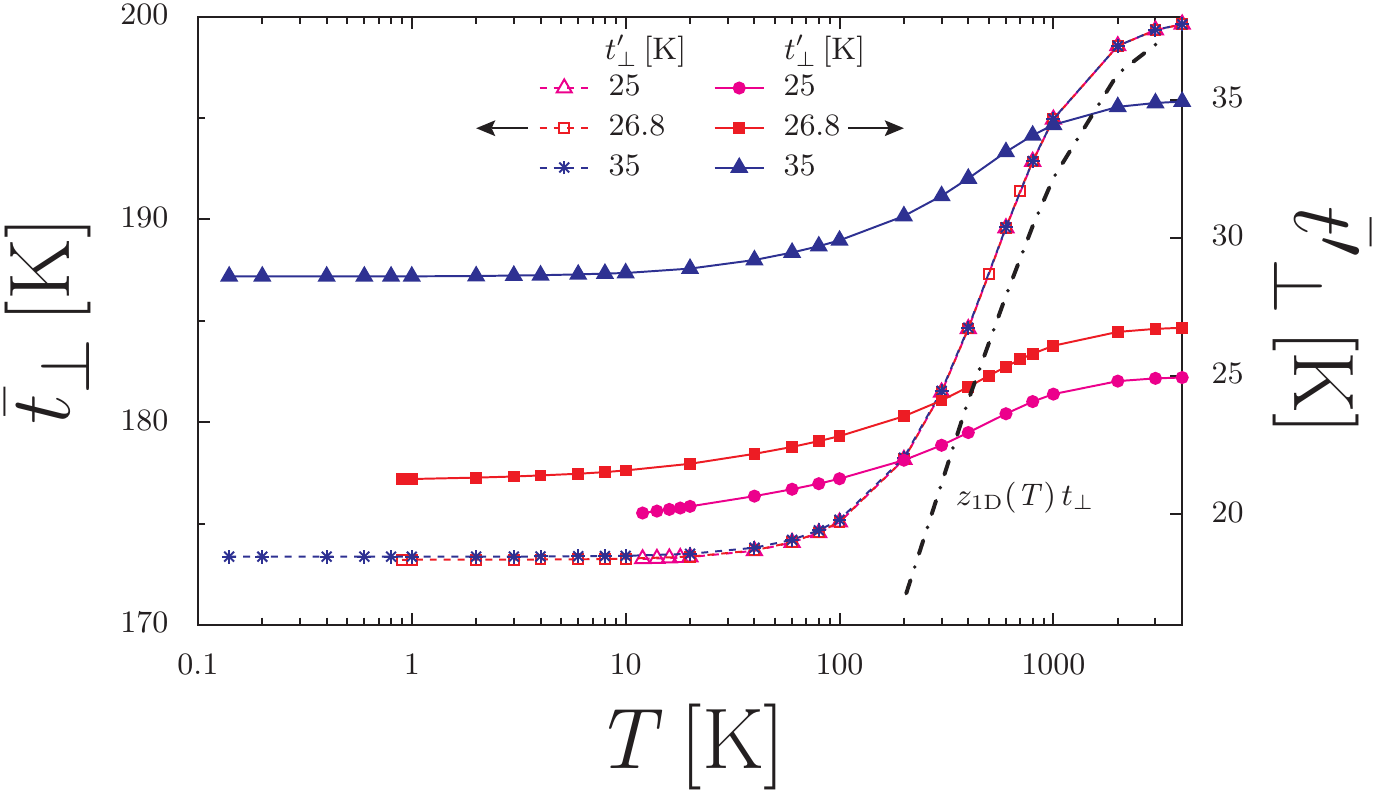}
 \caption{(Color on line) Renormalized interchain hopping $\bar{t}_\perp$   and   antinesting term  $\bar{t}_\perp'$  by self-energy corrections as a function of temperature and for different bare values of $t_\perp'$. The dashed dot line corresponds to the result of extended scaling hypothesis for the renormalization of $t_\perp$.  \label{tbar}} 
 \end{figure}
 The temperature dependences of  $\bar{t}_\perp$ and  $\bar{t}_\perp'$ are shown in Fig.~\ref{tbar}  for different bare  $t_\perp'$ values. According to the Figure, one observes a  downward shift of $\bar{t}_\perp$,    taking   place above   $ T_X$ where it follows a decay of the form $\bar{t}_\perp \sim z_{1D}(T) t_\perp$. In    the 1D  temperature domain the amplitude of  $\bar{t}_\perp$ is therefore  governed by the reduction of the quasi-particle weight (dashed dotted line of Fig.~\ref{tbar}),  as predicted  long ago by the extended scaling hypothesis\cite{Bourbon84,Boies95}.  The downward renormalization, which is about 13\% of the bare $t_\perp$  for the model parameters used, is essentially $t_\perp'$ independent and does not correlate with $T_c$. The renormalization of $t_\perp$ is taking place for the most part above $T_X$ and is a function of the initial couplings.
 The present approach can describe continuously the whole temperature domain   including  the crossover toward the coherent transverse single particle motion below $T_X$ where the renormalization of $\bar{t}_\perp$ levels off (Fig.~\ref{tbar}); this occurs despite a quasi-particle weight that  still decreases below $T_X$ (Fig.~\ref{z}).
 
If we now turn to the reduction of the antinesting term $t_\perp'$, it is small  in the  1D temperature regime and  becomes significant below $T_X$ where the details of the Fermi surface become  coherent. The amplitude of the reduction attained  is about 15-20\% of the bare value  at very low temperature  and all $t_\perp'$.

\subsection{Retarded one-particle self-energy}
From the expression (\ref{G}) of the Green function, the  $\ell\to \infty$ Matsubara self-energy at   arbitrary $\omega_\nu$  on the FS  is given by  
\begin{equation} 
\begin{split}
\Sigma_p(\bar{k}_\perp)=  &\  i\omega_\nu [1-z(\bar{k}_\perp)] 
  - z_\perp(\bar{k}_\perp),
\end{split}
\end{equation}
Using a Pad\'e approximants procedure,\cite{Vidberg77} $\Sigma_p(k_\perp, i\omega_\nu)$ is  analytically continued   to   the retarded form $ \Sigma_p^{\rm R}(k_\perp,\omega) =\Sigma_p'(k_\perp, \omega) + i\Sigma_p''(k_\perp, \omega) $  on the real $\omega$ axis,   made up of the real ($\Sigma_p'$) and imaginary ($\Sigma_p''$) parts. The analytical continuation is carried out   for a discrete set of $N_\omega$ Matsubara frequencies, whose distribution is peaked around $\nu=0$ and spread with a decreasing density as $\nu$  grows to large values. To obtain $\Sigma_p^{\rm R}(k_\perp,\omega)$ over    the whole  frequency range,   extending up to $\pm {3\over 2} E_0 (=$9000K)   with respect to the  bare Fermi level at $\omega=0$, good convergence in the $\omega$ dependence is achieved for $N_\omega \sim 20$ or so. However, when focussing on the behavior of spectral quantities for small energy intervals centered on $\omega=0$, the most accurate results are obtained   at smaller $N_\omega \sim 10$ close to $\nu =0$.

\subsubsection{Real and Imaginary parts}
\begin{figure}
 \includegraphics[width=9.0cm]{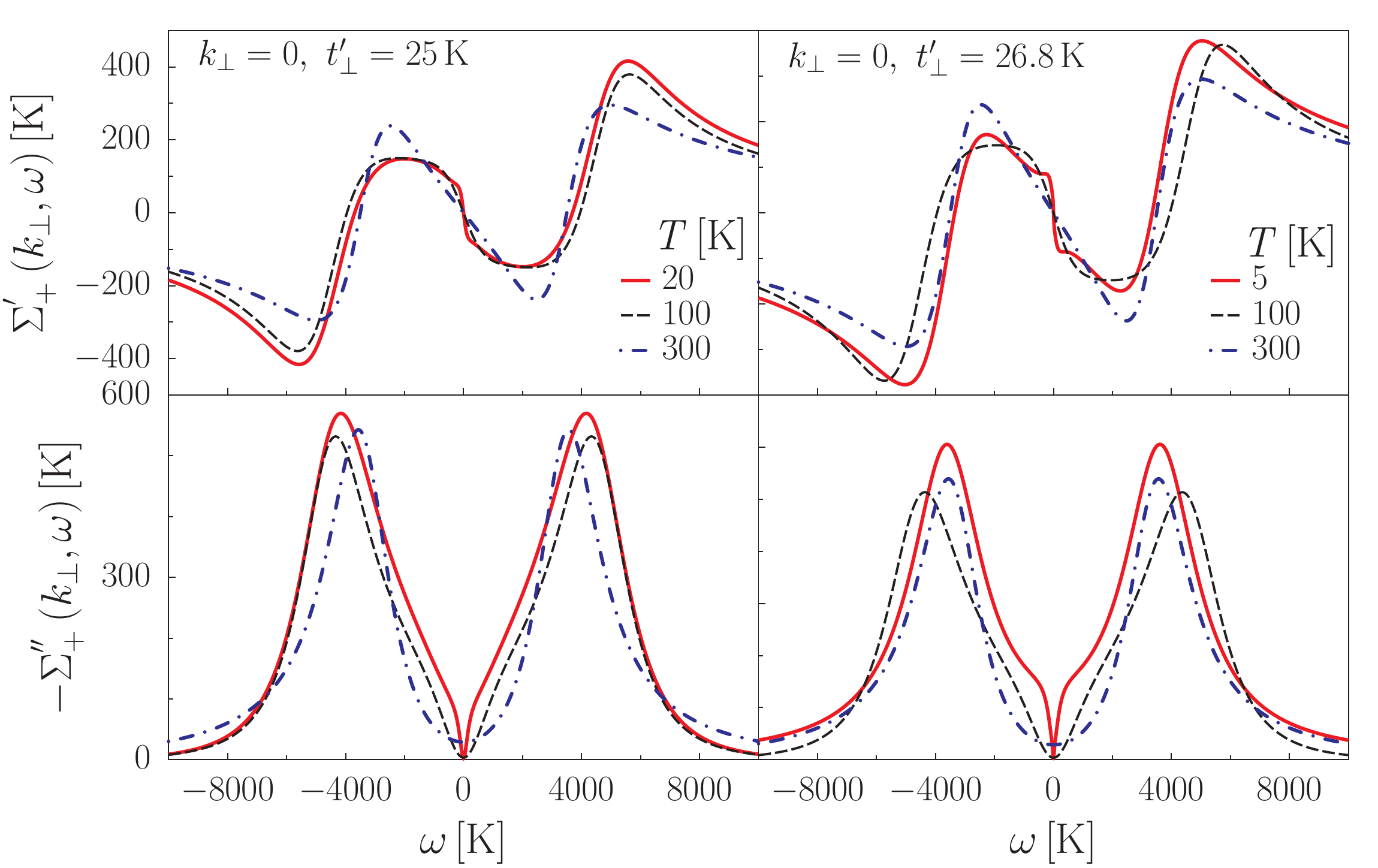}  \caption{ (Color on line) Typical  frequency dependence of the real ($\Sigma'$) and imaginary ($\Sigma''$)  parts of the one-particle self-energy at  different temperatures   in the SDW ($t_\perp'=25$~K; $T=20, 100, 300$~K, left)  and SC ($t_\perp'=26.8$~K; $T=5, 100, 300$~K, right) parts of the phase diagram in Fig.~\ref{Phases}.\label{RIm}} 
 \end{figure}

Typical results for   $\Sigma'_+(k_\perp=0,\omega)  $ and $\Sigma_+''(k_\perp=0,\omega)$  over the whole frequency  range   are shown in Fig.~\ref{RIm} in the SDW and SCd sectors of the phase diagram. 
 The general features shown correspond to those expected for a metal with single particle excitations. The real part  $\Sigma_+' $ for example, presents  two minima and maxima, and  crosses the  
Fermi level linearly  with a negative slope,  as  $\omega \to 0$. To the resonant  structure of $\Sigma_+'$  at low frequency corresponds  a minimum in  minus the   imaginary part, $-\Sigma_+''$   at $\omega=0$, as shown in Fig.~\ref{RIm}.

 The slope in $\Sigma_+' $ at $\omega\to 0$  enters in the determination of  the quasi-particle weight  by  the Fermi liquid expression 
\begin{equation}
\label{zslope}
z({k_\perp})=[1-\big(\partial \Sigma_+'(k_\perp,\omega)/\partial \omega\big)_{\omega \to 0}]^{-1}.
\end{equation}   
Thus the accuracy of the analytical continuation can be verified by comparing (\ref{zslope})  to the result   discussed previously following    the direct  evaluation  of Eq.~(\ref{zeq}).  In all cases studied,  both determinations match to a very good degree of accuracy (one percent or less of discrepancy at arbitrary $T$).

In the 1D temperature domain   at \hbox{$T = 300~{\rm K} >T_X$}  and for both values of $t_\perp'$, the linear part of $\Sigma_+'$ that crosses the origin  spreads   over the large   energy  interval, $\delta\omega\sim 2\pi T$ (Fig.~\ref{RIm}). In this temperature range, the quasi-particle weight  essentially  coincides with  its 1D limit, $z_{\rm 1D}$, which  varies as a power law in temperature -- consistently with the results of Sec.~\ref{zSec} (Fig.~\ref{z}). To the $\omega$-linear behavior within $\delta \omega$ in $\Sigma_+'$ corresponds a rounded minimum   centered at $\omega=0$  for $-\Sigma_+''$, well described by  the  quadratic    dependence, $ -\Sigma_+'' \propto \omega^2$ --  known to hold for a Fermi liquid \cite{Luttinger61}.  So despite the Luttinger liquid power law reduction of $z_{1D}$ in temperature,   Fermi liquid type corrections  in    frequency   remain within  a finite energy interval fixed by the temperature. Similar Fermi liquid  corrections in frequency   around the FS can be found in  the exact   spectral quantities of the 1D Tomanaga-Luttinger model  at finite $T$.\cite{Nakamura97b,Nakamura97}

As the temperature is lowered below $T_X$,  the width $\delta \omega$ where the linear part of $\Sigma_+'$   takes place   narrows markedly. The slope at $\omega \to 0$  becomes steeper, which in accordance to (\ref{zslope}) is indicative of a decreasing quasi-particle weight, as  obtained from  (\ref{zeq}) and   Fig.~\ref{z}. The narrow width of the $\omega$-linear variation in $\Sigma_+'$ is still correlated   with  the  quadratic frequency dependence in $-\Sigma_+''$  (Fig.~\ref{RIm}). Outside $\delta \omega$, however, the dip structure in $-\Sigma_+''$ sharpens considerably and is no longer quadratic in frequency.   
  \begin{figure}
 \includegraphics[width=8.0cm]{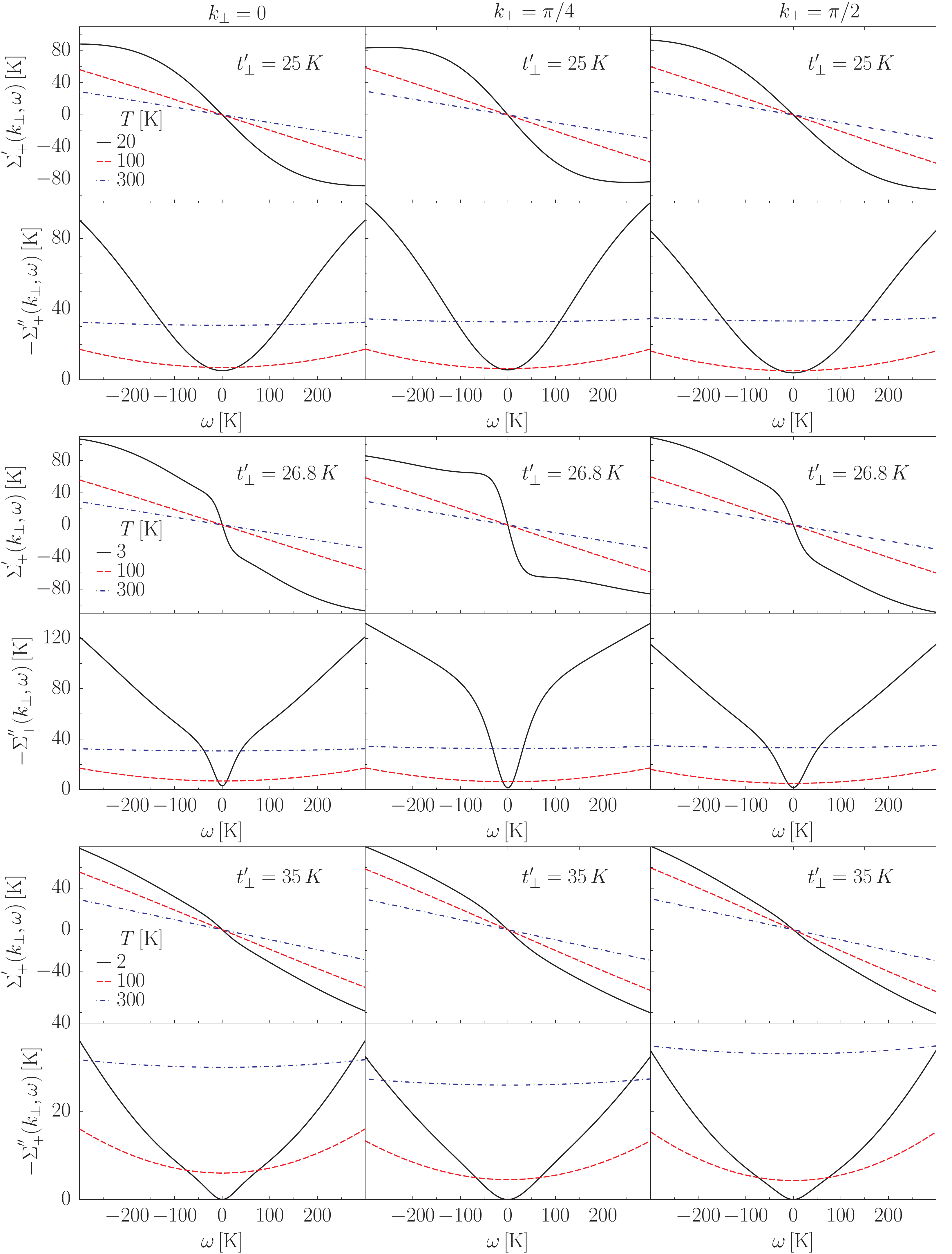}  \caption{ (Color line) Low frequency dependence of the real ($\Sigma_+'$) and imaginary ($\Sigma_+''$) parts of the one-particle self-energy at   different   $t_\perp' $ and regions of the Fermi surface. \label{Sw}} 
 \end{figure}

The modification of the quasi-particle resonance at low temperature can be examined in more details by focussing on   low frequency  dependence of  $\Sigma_+'$ and $-\Sigma_+''$,  in the range  $|\omega|\lesssim \pi T_X$ and  for different $t_\perp'$, as shown in  Fig.~\ref{Sw}. According to the Figure, we verify  that for all $t_\perp'$, the quasi-particle resonance shape in  $\Sigma_+'$ modifies and crosses  linearly  the   origin    with a  slope that becomes steeper as $T$ is lowered  within $\delta \omega$. The  slope along with the  amplitude of the resonance in $\Sigma_+'$ are the most pronounced close to $t_\perp'^*$ and  fall  off as $t_\perp'$ moves far apart from the SDW-SCd juncture, that is  when $T_c$ becomes small.

Well below $T_X$, the modification of $\omega$-linear term in $\Sigma_+'$      co-occurs with the development of a narrow dip in  $-\Sigma_+''$ that  invariably shows a    quadratic  dependence within $\delta \omega$.  The expression for the real part,    
\begin{equation}
\Sigma'_+(k_\perp,\omega) \approx -\lambda(k_\perp,T)\omega, \ \ \ \ \ (|\omega| <\delta \omega/2),
\end{equation}
though linear in $\omega$ at small frequency, as in a Fermi liquid,  has a momentum dependent linear coefficient  $\lambda(k_\perp,T)$ related to the anomalous decay of quasi-particle weight  ($\lambda = z^{-1}-1$).  As we have seen previously, $\lambda(k_\perp,T)=g(k_\perp) \ln T_{\rm CW}/T$  in the CW regime of the SCd sector and  is well approximated by  marginal Fermi liquid logarithmic corrections (Fig.~\ref{zmfl}).

Well outside the range of frequency fixed by $\delta \omega$ at low temperature,   deviations to linearity in $\Sigma_+'$ yield  as much departures from quadratic variations in $-\Sigma_+''$ (Fig.~\ref{Sw}). The latter are rather  well described by a  linear frequency 
dependence of the form, $-\Sigma_+'' (k_\perp,\omega) \approx \alpha_0(k_\perp) +  \alpha(k_\perp)|\omega| $   \cite{Sedeki10}. The extrapolation shows a finite intercept as $\omega\to 0$ ($\alpha_0\ne 0$), due to the   $\omega^2$ dependence within $\delta \omega$. The size of the linear factor $\alpha(k_\perp)$  falls off as  $t_\perp'$ is increasing  or $T_c$ decreasing. 

Regarding the $k_\perp$ dependence exhibited by $\Sigma_+'$ and   $\Sigma_+''$,  it is  present albeit not large. The steepest slope at zero frequency is found at $k_\perp =\pi/4$ (and  similarly  at $-\pi/4$, $\pm 3\pi/4$), in agreement with the minima  of $z(k_\perp)$ found in Fig.~\ref{zk} and that  are most  pronounced  close to $t_\perp'^*$.

\subsubsection{Spectral weight}
\begin{figure}
 \includegraphics[width=8.5cm]{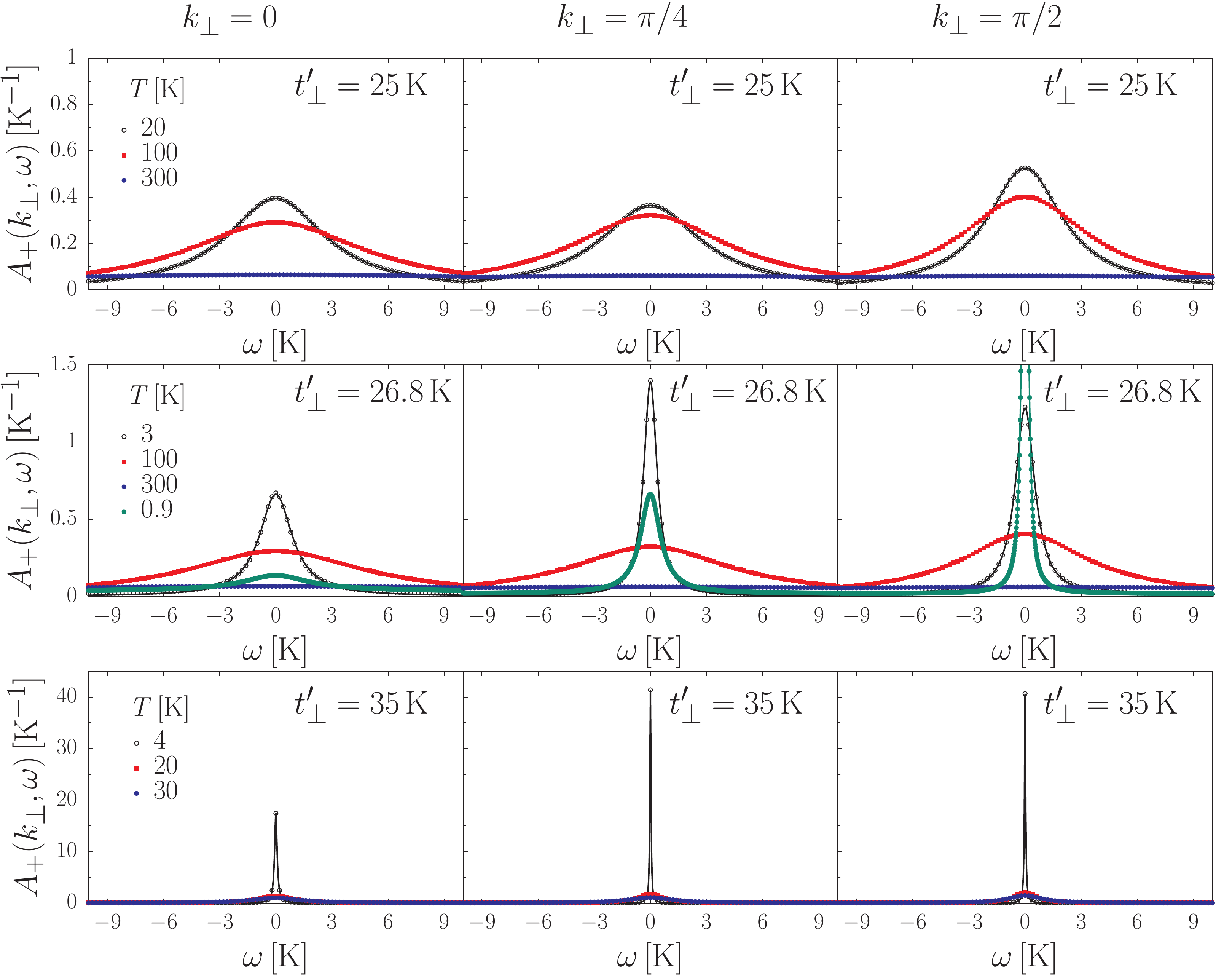}  \caption{(Color on line) Low frequency dependence of the single particle spectral weight  at  different $t_\perp' $ and  regions of the Fermi surface.\label{A}} 
 \end{figure}

The nature of single particle excitations on the FS as a function of the antinesting parameter $t_\perp'$ or $T_c$ can be further sharpened  by looking at the spectral weight $A_p({\bf k}_{F,p},\omega)=-2{\rm Im}\,G^{\rm R}_p({\bf k}_{F,p},\omega) $,   defined with the aid of the retarded  Green function. For $p=+$, for instance, one has
 \begin{figure}
 \includegraphics[width=8.0cm]{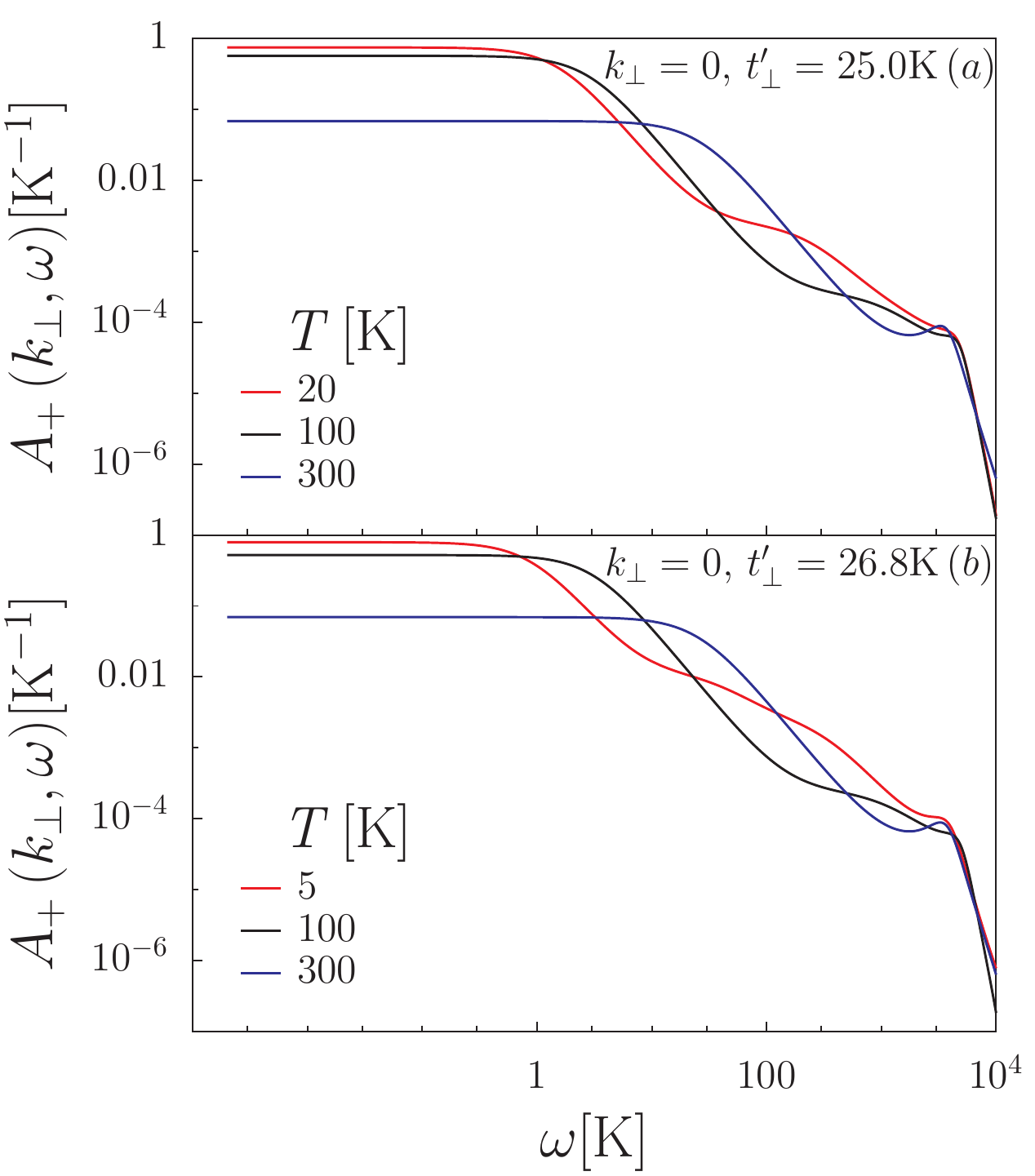}  \caption{(Color on line) Spectral weight at $k_\perp=0$ on the Fermi surface versus frequency (log-log scale) in the metallic domain of the SDW ($t_\perp' =25$K) and SCd ($t_\perp'=26.8$K) sectors of the phase diagram.\label{Aw}} 
 \end{figure}
\begin{equation}
A_+({ k}_\perp,\omega) =   { -2    \Sigma_+''({ k}_\perp,\omega)\over [\omega -   \Sigma_+'({k}_\perp,\omega)]^2 + [  \Sigma_+''({ k}_\perp,\omega)]^2}.
\label{Akw}
\end{equation}
 Typical normal state profiles of $A_+({ k}_\perp=0,\omega)$ over the whole  range of frequencies  are  shown in Fig.~\ref{A} at different  $T$ on each side of  $t_\perp'^*$. 
 
 For both $t_\perp'$, one finds  a growing  peak at $\omega=0$ as $T$ decreases  
 while its   width in frequency is reduced. 
 From Fig.~\ref{Aw}, one also   notices  that with decreasing $T$ the spectral weight   is redistributed    over a sizable part of the frequency range.  Although the redistribution is not  large in amplitude, the shift of spectral density toward high frequencies reflects the influence of interactions in the  build-up  of an incoherent part    of the spectrum in the mid-frequency range, which is   concomitant with the lost  of   quasi-particle weight $z(k_\perp)$ on the FS   associated to the coherent counterpart  (Fig.~\ref{z}). The redistribution is consistent with the sum rule, 
$$
(2\pi)^{-1}\int_{-\infty}^{+\infty} A_+({ k}_\perp,\omega) d\omega=1,
$$ 
which is  accurately satisfied   as a function of  $k_\perp$, for arbitrary $T$ and $t'_\perp$.

In Fig.~\ref{A}, we present a close-up view of the coherent component    of the spectral weight   at small frequency    for different $t_\perp'$, $T$, and $k_\perp$. We see that   on  the SDW side at low temperature (top panel of Fig.~\ref{A}),  the height of the central peak in $A_+({ k}_\perp,\omega)$ is  maximum at $k_\perp=\pi/2$, whereas the minimum is at   $k_\perp =\pi/4$. With roughly the same width $\Delta \omega$ in frequency, these peaks correlate  with the minimum (maximum) found     in the  quasi-particle weight for the best (worst) nesting conditions (Fig.~\ref{zk}), through the  Fermi liquid  relation 
$$
z(k_\perp) = (2\pi)^{-1} \int_{\Delta \omega} A_+(k_\perp,\omega)d\omega. 
$$ 
However, as we enter   the  low temperature metallic phase  of the SCd sector of the phase diagram at  intermediate and large $t_\perp'$, there is a notable change in the positions of   maxima and minima  of $A_+(k_\perp,\omega)$     along the Fermi surface. Under the  control  of the SCd fixed point in the CW domain,   the quasi-particle peak  becomes  according to  Fig.~\ref{A}  maximum  at $k_\perp=\pi/4$ (also  at $-\pi/4$, $\pm 3\pi/4$)  and minimum at $k_\perp=0$ (also at $\pm \pi$). The  width of the peaks and the tails, though   remaining   sizable at very low temperature,  progressively narrow  while the height raise    as $t_\perp'$ moves away from $ t_\perp'^*$. It is only in the limit of large $t_\perp'$ or very small $T_c$ that the peaks evolve toward the formation of a Dirac like structure at $\omega=0$ (Fig.~\ref{A}, bottom panel).  Put together with the very weak  temperature decay of the quasi-particle weight  (Figs.~\ref{z} and \ref{zmfl}) and the dominant  quadratic temperature dependence of the scattering rate (see Sec.~\ref{Rate}), this limit is symptomatic of the emergence of a Fermi liquid. A similar Fermi liquid recovery for  quasi-particles on the SCd side is found at the nodal points, $k_\perp\pm \pi/2$,   as $T\to T_c$, while for the rest antinodal region, the spectral weight is suppressed by critical fluctuations preceding  the opening of the d-wave  gap (Fig.~\ref{A}, middle).

\subsubsection{Electron-electron scattering rate}
\label{Rate}
\begin{figure}
   \includegraphics[width=5.0cm]{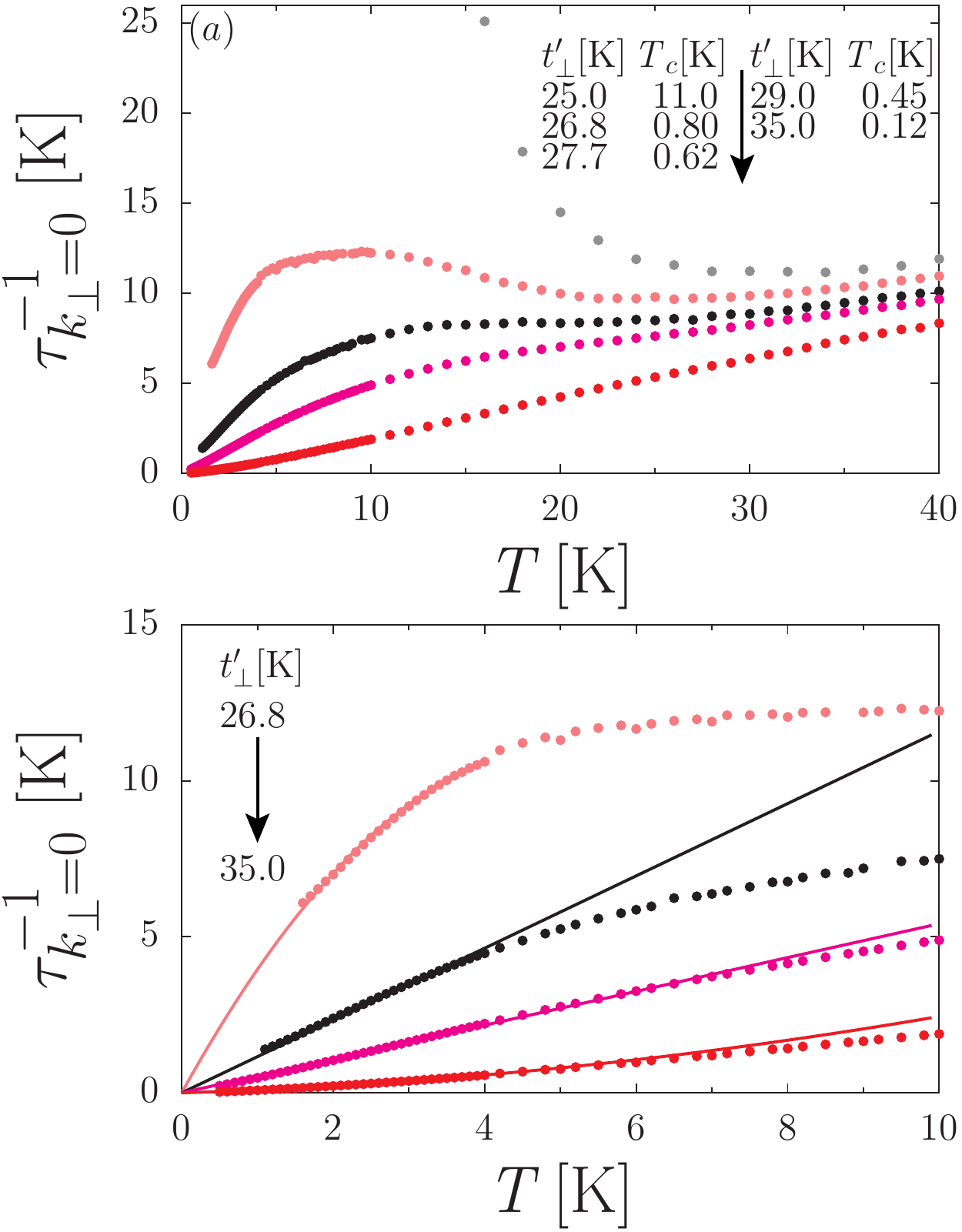}  \caption{(Color on line) (a) : Temperature dependence of  the   electron-electron scattering rate  at $k_\perp=0$ for different  $t_\perp'$ in the SDW and SCd regions of the phase diagram; (b) : Curie-Weiss temperature region in the SCd domain where the continuous lines correspond to a fit to Eq.~(\ref{tauk}). \label{Rate1}} 
 \end{figure}

An important quantity giving key information on the impact of correlations on single particle properties on the FS is the electron-electron scattering rate. This quantity is  defined by the imaginary part of the self-energy at $\omega=0$, 
\begin{equation}
\tau_{k_\perp}^{-1}=-2\Sigma''_+(k_\perp,0).
\end{equation}
In Fig.~\ref{Rate1}, the $ \tau_{k_\perp}^{-1}$ temperature dependence   at   $k_\perp=0$ is shown for $t_\perp'$  in  the SDW ($t_\perp'=25$K)   and  SCd ($t_\perp'=26.8-35$K)  regions  of the phase diagram.

 For SDW, $\tau_{k_\perp=0}^{-1}$ decreases from high temperature and   ultimately shows an upturn at low temperature at the approach of the critical domain of $T_{\rm SDW}$ ($\simeq$ 12K), where couplings become  strong and short-range SDW fluctuations   act as a singular source of scattering for quasi-particles. 
On the SCd  side,$\tau_{k_\perp=0}^{-1}$ slowly decreases with decreasing temperature,  then   crosses over toward  the CW regime below 10K (Fig.~\ref{Phases}), where   $\tau_{k_\perp=0}^{-1}$ falls off toward zero. The  drop  corresponds to the   sharpening of the coherent peak in the spectral weight for  the same temperature interval (Fig.~\ref{A}); it  continues down to the onset of critical SCd domain close  to $T_c$ where a sudden rise in $\tau_{k_\perp=0}^{-1}$ occurs (not shown in the Figure).     

\begin{figure}
   \includegraphics[width=5.0cm]{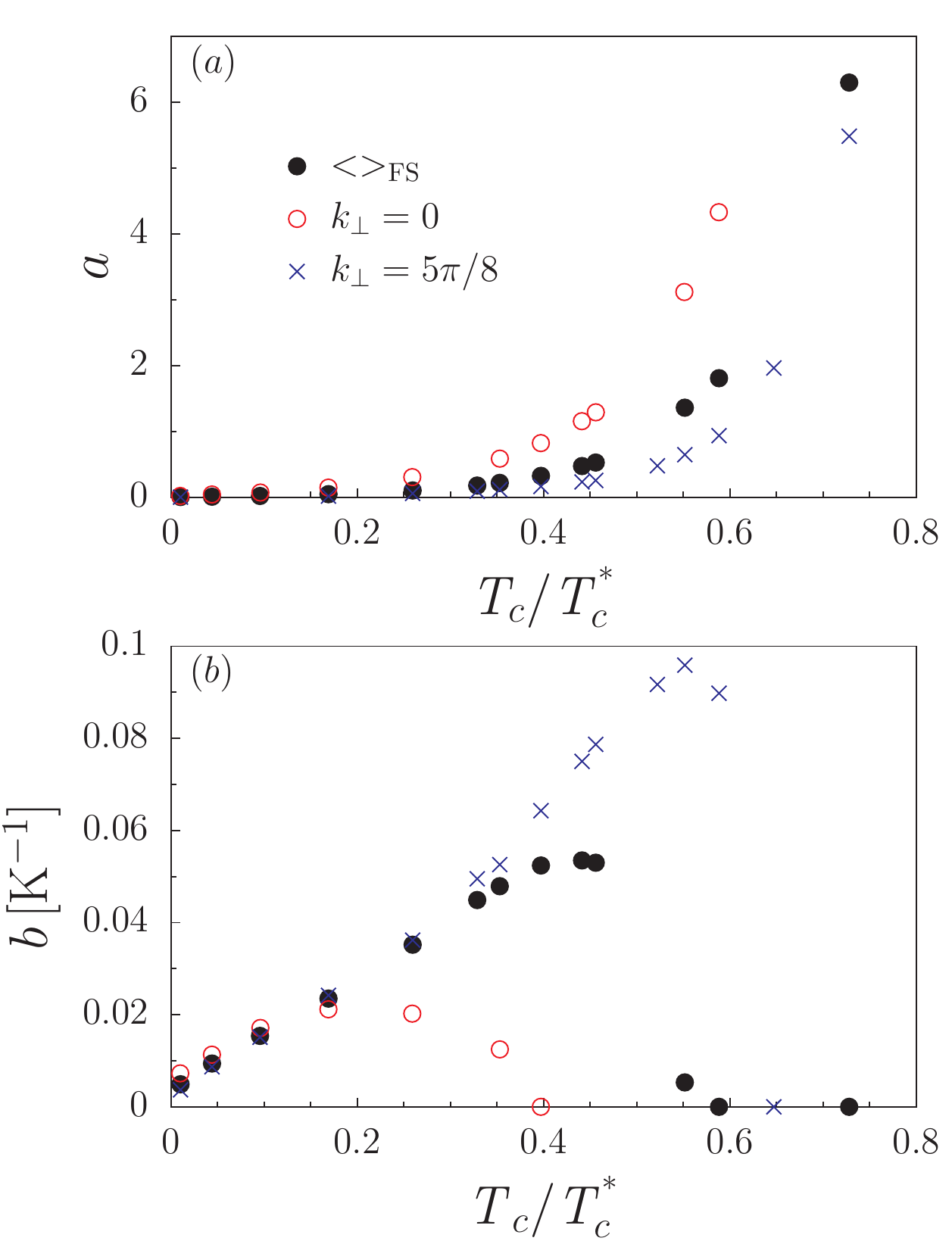}  \caption{(Color on line) Evolution of the  linear (a) and quadratic (b) coefficients of the scattering rate as a function of the ratio  $T_c/T_c^*$ for $k_\perp=0$ (open circles) and $k_\perp=5\pi/8$ (crosses). The full circles correspond to the averages $a$ and $b$ over the Fermi surface (Eq.~\ref{Poly}). \label{ab} }
 \end{figure}

Within the CW temperature domain, the scattering rate  is well described by a polynomial temperature dependence of the form
\begin{equation}
\label{tauk}
 \tau_{k_\perp}^{-1} \approx  a(k_\perp)T + b(k_\perp)T^2,
\end{equation}
which comprises an anomalous linear component and quadratic Fermi liquid term. The fit of the $k_\perp=0$ results to the above expression in the normal phase interval $ ]T_c,  4~{\rm K}]$ of the CW domain  reveals a pronounced linear component  with the $a(k_\perp=0)$ coefficient that decreases with $t_\perp'$ or the decay of $T_c$ (Figs.~\ref{Rate1}-b and \ref{ab}-a). The Fermi liquid component is only emerging well above $t_\perp'^*$ where $\tau_{k_\perp=0}^{-1}$ begins to show a positive curvature. In the limit of large $t_\perp'$ or small $T_c$, the anomalous part  is vanishingly small and only the Fermi liquid contribution remains, in concordance with the  virtually temperature independent quasi-particle weight (Figs.~\ref{z}-\ref{zmfl}) and the delta-like peak  in the spectral weight (Fig.~\ref{A}).

\begin{figure}
   \includegraphics[width=5.0cm]{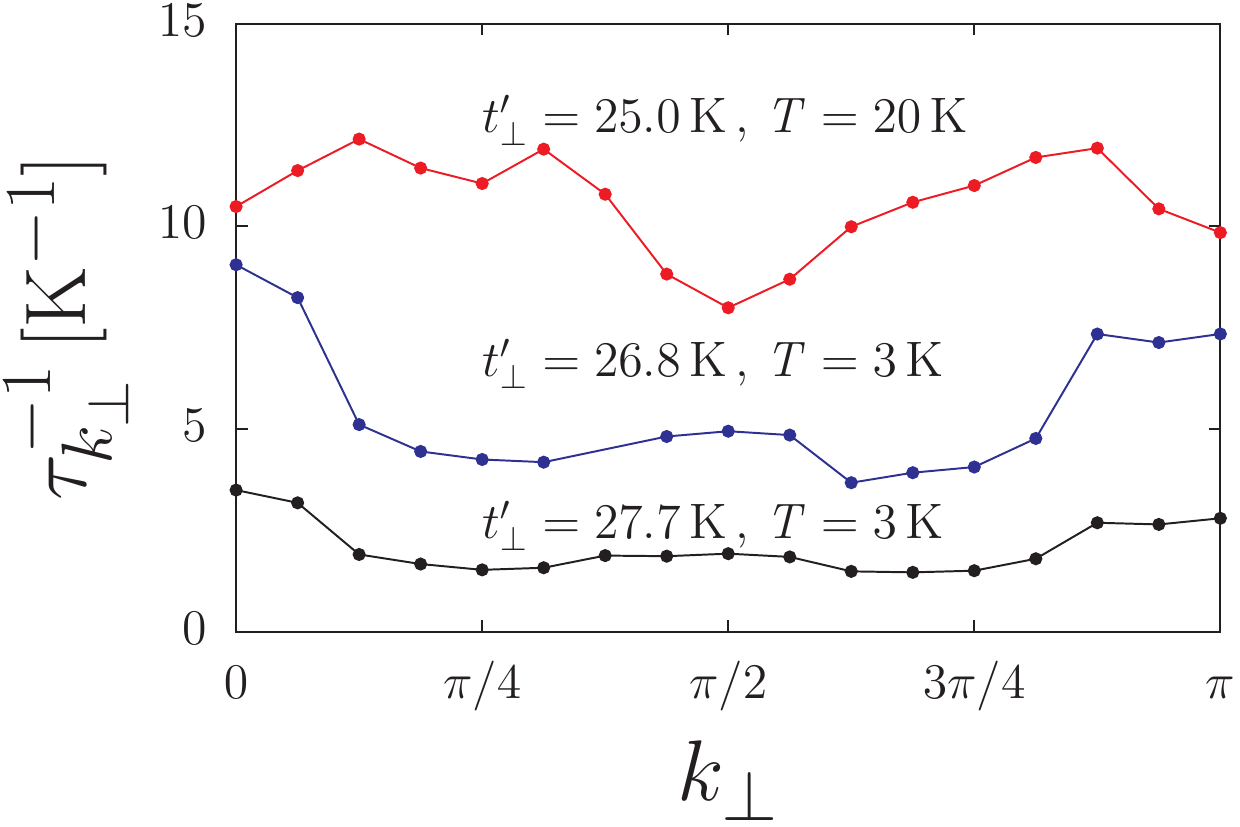}  \caption{(Color on line) Momentum dependence of the  quasi-particle scattering rate on the Fermi surface in the SDW ( $t_\perp'=25$K) and SCd  ($t_\perp'=26.8$ and 27.7K). \label{Ratek}} 
 \end{figure}

It is instructive to look at the  $k_\perp$ dependence of the scattering rate  for both sectors of the phase diagram. As shown    in Fig.~\ref{Ratek} for the  SDW metallic phase, there is a  correspondence between the best (worst) nested regions of the FS at $k_\perp\simeq \pm \pi/4$ or $\pm 3\pi/4$ ($\pm \pi/2$) where  $\tau_{k_\perp}^{-1}$ is maximum (minimum) and those where  $z(k_\perp)$ is minimum (maximum) (Fig.~\ref{zk}). This is consistent with  the usual picture of `warmer' regions of the FS governed  by stronger  scattering on SDW fluctuations and better  nesting conditions. 
By   contrast, this correspondence no longer holds  in the CW metallic regime on the SCd side: the best nested regions are where the scattering rate is the lowest: the maxima being shifted  at $k_\perp =0$  and $\pm \pi$ (Fig.~\ref{Ratek}).
Thus,  although SDW correlations remain the strongest in the metallic state, that is  $\chi_{\rm SDW} \gg \chi_{\rm SCd}$ at all $T$ (except very close to $T_c$), the SDW    source of scattering is clearly affected by the presence of Cooper pairing. This shift in the $\tau_{k_\perp}^{-1}$ maxima is consistent with the one encountered for the spectral weight as $t_\perp'$ crosses over to the SCd side (Fig.~\ref{A}).

\begin{figure}
 \includegraphics[width=5.0cm]{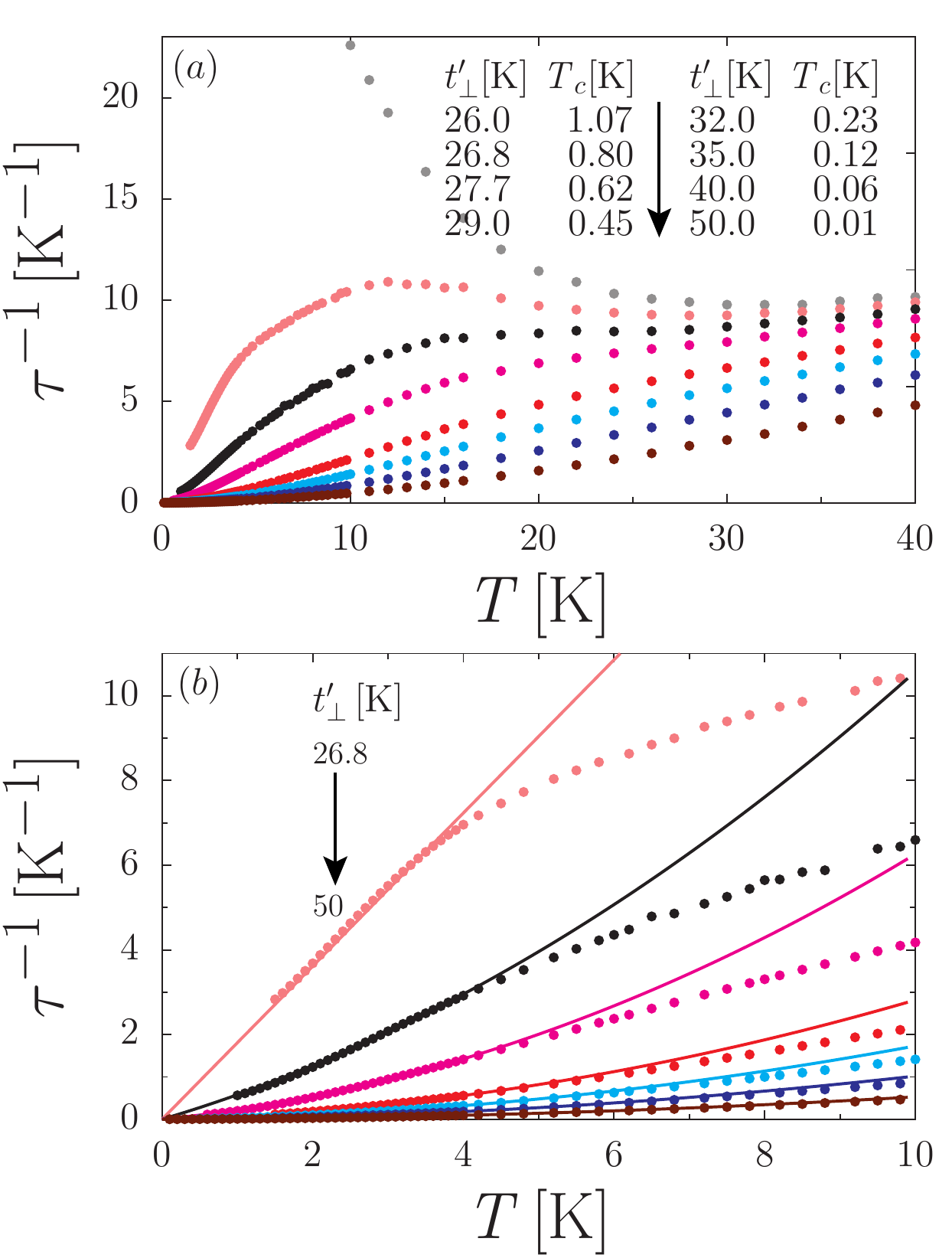}  \caption{(Color on line) (a) : Temperature dependence of   the electron-electron scattering rate averaged on the Fermi surface for different  $t_\perp'$ in the superconducting sector of the phase diagram; (b):  polynomial fit (continuous lines) of the temperature dependence of the mean scattering rate (Eq.~\ref{Poly}).  \label{RateAv}} 
 \end{figure}

 In spite  of the $k_\perp$ anisotropy   shown  by $\tau^{-1}_{k_\perp}$  in Fig.~\ref{Ratek},  the    temperature dependence  along the Fermi surface turns out to be  similar as a function of $k_\perp$,  only the amplitude  of the $a(k_\perp)$ and $b(k_\perp)$ of the polynomial form (\ref{tauk}) varies. It is therefore useful to  concentrate on   the averaged scattering rate, $\tau^{-1}\equiv \langle \tau_{k_\perp}^{-1}\rangle_{\rm FS}$, over the FS.  The  average scattering rate is shown  in   Fig.~\ref{RateAv}   at different $ t_\perp'>t_\perp'^*$  in  the superconducting domain.

Consider  first the  SCd region   close to the juncture with SDW,  near the critical $t_\perp'^*$. As shown in Fig.~\ref{RateAv}, the scattering rate decreases  from    high  temperature, but ultimately  shows an upturn with lowering $T$, which is typical of a SDW type of behavior (see Fig.~\ref{Rate1}-a).  In spite of an SCd instability toward superconductivity at $T_c \sim 1$K,  SDW correlations are  so large that  the scattering becomes critical or   `insulating' like. This strong coupling behavior accords with the marked suppression of quasi-particle weight  discussed  for the same $t_\perp'$ conditions  in Fig.~\ref{zmfl}. Strong coupling effects close to $t_\perp'^*$ may indicate a pseudogap type  behavior and signal  the limitations  of weak coupling one-loop  calculations for scattering amplitudes. 

As $t_\perp'$ is gradually tuned  away from $t_\perp'^*$, the  $\tau^{-1}$ variation modifies  around 10K.  Following the example of the $k_\perp=0$ case in Fig.~\ref{Rate1},  as $T\to 0$,  $\tau^{-1}$ extrapolates toward zero in  the CW domain. Metallic one-particle  states then sharpen with a steadily growing lifetime  down to the onset of critical SCd fluctuations (Fig.~\ref{RateAv}).  The corresponding temperature dependence of $\tau^{-1}$   is rather  well  described by   the  averaged   polynomial form of (\ref{tauk})\cite{PowerFit}, 
\begin{equation}
\label{Poly}
  \tau^{-1}  \approx  a T +   b  T^2,
\end{equation} 
where  $a=\langle a(k_\perp)\rangle_{\rm FS}$ and $b=\langle b(k_\perp)\rangle_{\rm FS}$ are the     coefficients averaged over the FS.   Therefore both a Fermi liquid and an anomalous  linear term superimpose in $\tau^{-1}$, as   shown by the continuous lines of Fig.~\ref{RateAv}-b obtained from a fit of (\ref{Poly}) to the $\tau^{-1}$ results in the same CW   temperature interval, $]T_c, 4{\rm K}]$,     used  previously for $k_\perp=0$. 

Not too far from  the boundary,  there is a finite $t_\perp'$ region where the  quadratic part vanishes and $\tau^{-1}$ is uniquely $T$-linear. By averaging the scattering rate, this region  is much  than for $k_\perp=0$ (Fig.~\ref{ab}-b).

A few remarks are in order concerning the origin of a  $T$-linear  component in $\tau^{-1}$.  It can  be seen as the result  of quasi-particles  scattering on SDW correlations whose spectral density    peaked at frequency lower or of the order of temperature, which would be compatible with the sizable  amplitude of $\tilde{\chi}_{\rm SDW}$ or large SDW correlation length $\xi$ (Fig.~\ref{Phases}). In these conditions,  electron scattering on  spin fluctuations  in two dimensions \cite{Vilk97,Abanov03},  is known to yield $\tau^{-1} \sim T $ when $\xi$ is weakly temperature  dependent, which is here  realized  for  $T\ll \Theta$. An alternative but related interpretation   of  the impact of vertex parts on  self-energy  diagrams  of Fig.~\ref{flow} is provided by  the phenomenology of the marginal Fermi liquid. In the latter scheme,    electron scattering   on dispersionless bosons  formed by particle-hole pair excitations      also  yields  $T$-linear scattering rate.\cite{Varma89}    Such an interpretation would be compatible   with the temperature decay of the quasi-particle weight,  which,  as we have seen previously, follows  the marginal Fermi liquid behavior in the same temperature domain of the normal phase (Fig.~\ref{zmfl}).   

As $t_\perp'$ is tuned upward  and $T_c$ decreases,  there is a  threshold in $t_\perp'$ above which the variation of $\tau^{-1} $ begins to develop 
a finite curvature   well described by a Fermi liquid $bT^2$ term (Fig.~\ref{RateAv}). According to Fig.~\ref{ab}, the quadratic coefficient is first growing rapidly from zero, reaches a maximum and ultimately   decreases  as one moves away from the SDW-SCd juncture and $T_c$ drops along the $t_\perp'$ axis; this non monotonic trace is likely to result  from  the interplay between the    stiffening of the spin fluctuations spectrum (increasing $\Theta$, Fig.~\ref{Phases}-a) and their decreasing  amplitude. 

As shown in  Fig.~\ref{ab}, the emergence  of a Fermi liquid coefficient $b$ co-occurs with  the weakening of the linear component as $t_\perp'$ increases or $T_c$ decreases. Remarkably, however, the  $T$-linear term persists over a large interval of $t_\perp'$ where  $T_c$ takes  appreciable values (Fig.~\ref{ab}). This `correlation' between $a$ and $T_c$ demarcates from what is usually expected from a quantum critical point driven by the SDW channel of fluctuations alone. In the latter situation, the influence of superconducting pairing is absent and the linear component of the scattering rate is ordinarily confined within a   cone  of quantum criticality    close to the critical value of the tuning parameter \cite{Moriya90} -- though recent calculations show some  departure   from this scheme within the framework of the single SDW fluctuation channel\cite{Bergeron11}. The presence of  Cooper-Peierls pairing  mixing  is here responsible for the expanse of  the  $T$-linear component   along the $t_\perp'$ axis. The possibility in the RG framework to switch off all Cooper pairing contributions in the  scattering amplitudes (\ref{gn})  allows a direct confirmation   of this mutual influence. When all Cooper loops in (\ref{gn}) are put to zero   and the Peierls channel is singled out, we   verify that linearity is indeed confined to a narrow region above $t_\perp'^*$, and  is further followed by the prompt emergence of a  Fermi liquid $bT^2$ term, with a $b$ coefficient   an order of magnitude  larger   than the one found when the two scattering channels are present (Fig.~\ref{ab}).  

\begin{figure}
 \includegraphics[width=5.0cm]{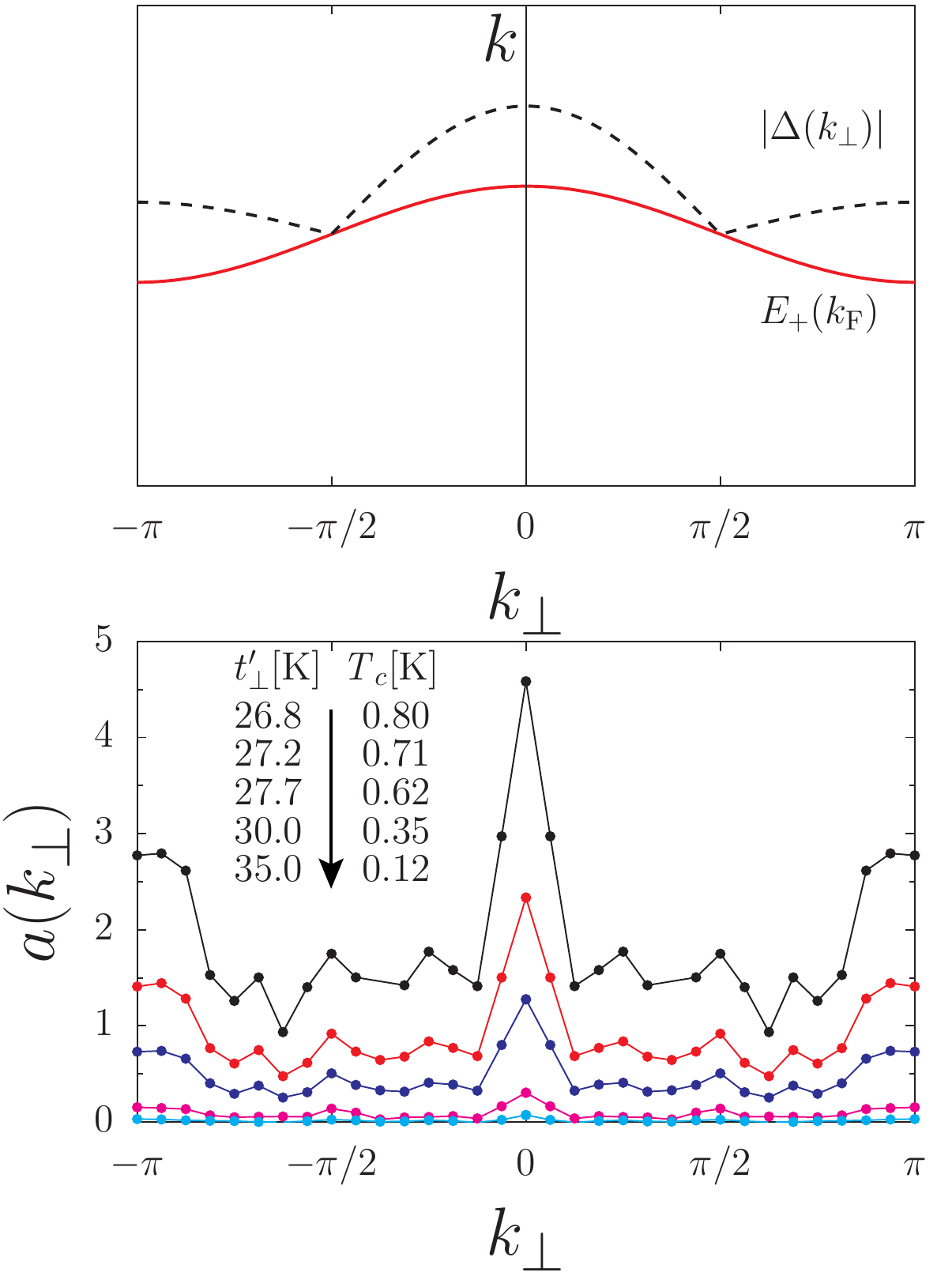}  \caption{ (Color on line)  Not to scale variation of the d-wave superconducting  gap modulus,   $|\Delta(k_\perp)|$,    along the $+k_F$ Fermi  surface sheet (top); $T$-linear coefficient   anisotropy of the quasi-particle scattering rate along the Fermi surface and at different  $t_\perp'$ (bottom).    \label{akp}} 
 \end{figure}

Another revealing fingerprint  of the contribution of  Cooper pairing  to the broad $t_\perp'$-range of  anomalous scattering rate can be  found in the anisotropy    of the linear-$T$ coefficient, $a(k_\perp)$,  along the FS.  As shown in Fig.~\ref{akp} for   different $t_\perp'$, a prominent maximum with secondary maxima are found at $k_\perp=0$ and $k_\perp= \pm \pi$ respectively. These positions are   clearly   where the superconducting d-wave gap,  $\Delta(k_\perp) = \Delta_0\cos k_\perp + {\cal O}(\cos 2k_\perp)$,  is expected to be maximum in amplitude  below $T_c$. The minima of $a(k_\perp)$ are found relatively close to the nodal region for the gap at $k_\perp\simeq \pm 5\pi/8 $ whose variation  with $T_c$ is also shown in Fig.~\ref{ab}.

This contrasts with the situation prevailing at either high temperature, namely  above the CW temperature domain, or in the metallic state of the SDW sector of the phase diagram  where $\tau_{k_\perp}^{-1}$ is peaked in the neighborhood of  the best nesting points at $k_\perp=\pm \pi/4$ or $\pm 3\pi/4$ (Fig.~\ref{Ratek}).   

\section{Discussion and concluding remarks}
In the normal phase associated to the SDW-SCd  sequence of instabilities of the quasi-1D electron gas, the above RG analysis brought out a number of anomalies in the single particle  quantities. For weakly dimerized chains  responsible for small Umklapp scattering,  and relatively weak electron-electron couplings, the temperature scales  for   instabilities are occurring far below the single particle crossover scale $T_X\sim t_\perp$, which expands the normal phase  at  lower temperature where single particle anomalies  become under the growing influence of  nesting deviations $t_\perp'$.

By tuning  nesting deviations along the $t_\perp'$ axis, the critical downturn  of $T_{\rm SDW}$ merges to  the  superconducting  instability line  at a much smaller $T_c$.   The  normal phase becomes therefore  enlarged    by the  Curie-Weiss interval of staggered spin fluctuations.   As shown by the temperature dependent SDW susceptibility, it extends up  to $T_{\rm CW}\sim 10$K, a scale   equal   to  many times $ T_c$. This temperature region  turns out   to be distinct in that    $T_{\rm CW}$  corresponds to  the  scale where the SDW fixed point becomes unstable  and    another  fixed point is emerging in  the SCd channel.  The evolution of the flow in the CW domain is  rather gradual, however, as shown by the continuous transformation  of interactions in momentum space.  Besides Umklapp, the RG flow shows that normal  scattering amplitudes at the SDW momentum transfer, ${\bf q}_0=(2k_F,\pi)$, are reinforced in the antinodal region of the FS, while  the  nodal  counterpart sees the scattering   weakened.\cite{Duprat01} Such an interplay is the fingerprint of a steady constructive interference between  Cooper (SCd) and Peierls (SDW) pairings  that takes place  in the CW  temperature interval.  It correlates in amplitude  with the distance from the critical $t_\perp'^*$ for the occurrence of superconductivity. When fed into the  two-loop one-particle self-energy, it yields the scaling of  number of NFL  quasi-particles properties  with $T_c$.

The fact that two simultaneous pairing processes interfere with and distort each other makes difficult a standard  description of low energy excitations of the normal phase in terms of   long wavelength order parameter fluctuations.\cite{Hertz76,Millis93}  In this regard,  it is of interest to compare the CW  situation with the one  that prevails in 1D systems  where such     entanglement of many-body processes  is at its maximum. These optimal conditions turn out to be  realized    here when $T\gg T_X$,   which are  well known to invalidate the Ginzburg-Landau-Wilson description of collective excitations in terms of order parameter fluctuations.  The proper collective excitations  in the 1D range are  essentially  given by  the     bosonic  modes of the Luttinger liquid,   which is quantum critical in character at all scales.\cite{Giamarchi04} In the low temperature regime, at $T \ll T_X$, the coherent warping of the FS and nesting deviations do attenuate interference, which becomes non uniform on the FS.  As stressed above, in spite of this distortion,   its presence   is maintained so that the mixing of electron-hole and electron-electron excitations   carries over  the flaws  of the usual  Ginzburg-Landau-Wilson  description of correlations   at lower temperature. When factored in, interference then  produces an expanded range of quantum criticality for the normal state and introduces an effective  correlation of    quasi-particle anomalies with $T_c$.

Extended  quantum criticality is  particularly well illustrated by the NFL  downturn of the quasi-particle weight $z(k_\perp)$ in the CW temperature domain. We have seen  that  for the whole non nodal part of the FS, $ z(k_\perp)$ is well described    by  temperature dependent logarithmic corrections  similar to those found in  the  marginal Fermi liquid theory. Since the  phenomenology of the marginal Fermi liquid is typically   quantum critical in character,\cite{Varma89,Vidhyadhiraja09}  the  persistent NFL  properties  shown    over  a broad range of $t_\perp'$ or $T_c$ is compatible with an expanded region of quantum criticality in the   phase diagram of the quasi-1D electron gas.

Another related  feature   shared  with this phenomenology   is supplied by the existence of a linear-$T$ component in the polynomial temperature dependence of the  scattering rate $\tau^{-1}_{k_\perp}$.   Following the example of     quasi-particle weight fraction,  the strength of  linear-$T$ scattering within  the same CW   region of the normal phase  also scales with $T_c$ along the $t_\perp'$ axis. 

These  findings on scattering rate  primarily  apply to the problem of anomalous electric transport  in the metallic state of (TMTSF)$_2 X$ salts. Within the  Curie-Weiss regime of SDW fluctuations   determined by NMR\cite{Brown08,Wu05,CreuzetRMN} the  resistivity data of  prototype compounds like $X= $PF$_6$ and ClO$_4$\cite{DoironLeyraud09,DoironLeyraud10,AubanSenzier11}, were shown to be well described by the polynomial form, $\Delta \rho\approx AT + BT^2$ (after   subtraction of the elastic   contribution from impurities).  From the critical pressure where superconductivity joins to antiferromagnetism in the phase diagram,\cite{DoironLeyraud09,Vuletic02} the linear coefficient $A$    shows a clear   scaling with $T_c$.  Under pressure, it becomes superimposed to  a  Fermi liquid  contribution that is quadratic in $T$. The latter    ultimately dominates the temperature dependence of resistivity in the high pressure  limit  where  the materials are no longer superconducting.

The above  features  of electrical resistivity in the Bechgaard salts adhere in many respects to the behavior  shown by the   scattering rate  as a function of $t_\perp'$ in the present model.
 In the presence of Umklapp scattering, the inelastic contribution to   resistivity in the absence of vertex corrections can be  written as $\Delta\rho = 4\pi   \langle \tau_{k_\perp}\rangle^{-1}_{\rm FS}/\omega_p^2$,  ($\omega_p$ is the plasma frequency), which is related to the inverse scattering time averaged over the FS -- the latter differing very little from $\langle \tau_{k_\perp}^{-1}\rangle_{\rm FS}$ considered in Sec.~\ref{Rate}\cite{DoironLeyraud10,Bourbon11}. Under these conditions, it follows that the problem of anomalous linear-$T$ resistivity and its correlation to $T_c$  for (TMTSF)$_2X$ can be viewed as the consequence of an expanded quantum critical regime for electrons that scatter off antiferromagnetic fluctuations. The  underlying interference of pairings may reveal   to be a key determinant in the mechanism that connects scattering  to $T_c$  in  the Bechgaard salts.

This brings us to consider whether  similar   ideas  can go beyond  the sole confines of low-$T_c$ materials and  apply to other unconventional superconductors.  In this respect iron-based pnictide superconductors are of  particular interest. There are indeed striking similarities shown by the SDW-SC sequence of ground states in many pnictides,   provided that   $T_c$ and $T_{\rm SDW}$ of the Bechgaard salts are magnified by a factor 20 or so.   The scaling of  both linear-$T$ resistivity,\cite{DoironLeyraud09a,DoironLeyraud09,Taillefer10}      and   Curie-Weiss regime of spin fluctuations with $T_c$, over a wide range of atomic substitution in pnictides,  \cite{Nakai10,Ning10} suggest that the mechanism described here for  extended criticality may be of more general  importance for the normal state of superconductors near a SDW instability.  Despite  a difference in their FS (more 2D, multiple sheets), the two basic active ingredients for extended criticality, namely, nesting and pairing, are known to be both  present and   interfering \cite{Wang09,Chubukov08,Thomale09}. 

The role of Cooper pairing in anomalous quantum criticality may be also relevant  to high-$T_c$  cuprates. In hole-type materials, for instance, common features with the organics can be found  in the overdoped region, namely where where the FS is large and coherent and the pseudo gap is absent.  
The hole-doped materials  are  among the first compounds  showing systematic  scaling of  anomalous resistivity      with $T_c$,\cite{DoironLeyraud09a,Taillefer10, Manako92,Cooper09} a temperature scale that can be two orders of magnitude larger than for  the Bechgaard salts. Nevertheless, as a function of hole concentration, resistivity evolves from perfectly $T$-linear,\cite{Cooper09,Daou09}  to the purely quadratic  dependence  of the Fermi liquid  once superconductivity is suppressed in the strongly overdoped region.\cite{Manako92,Nakamae03}  Results of functional RG on the 2D Hubbard model indicate that spin fluctuations which mix with  electron-electron scattering processes of large  momentum transfer can yield a linear component in the scattering rate  whose amplitude correlates with band filling and therefore $T_c$. \cite{Ossadnik08} In the framework of the two-particle self-consistent approach to the antiferromagnetic channel  of 2D Hubbard model,\cite{Vilk97}  the   calculation  of  resistivity,
     which takes into account  vertex corrections  and the extended range of spin fluctuation effects  in 2D, 
shows the presence of  anomalous temperature dependence for resistivity   over a sizable domain of carrier concentration.\cite{Bergeron11} 
   
     For  electron-doped  cuprates,\cite{Armitage10,Kyung2003,Hassan2008}    antiferromagnetism  borders on superconductivity as a function of electron concentration, in a  way  analogous  to the Bechgaard salts under pressure.  In these cuprates too, definite $T$-linear resistivity  is observed  close to optimal doping\cite{Fournier98}; it gradually weakens  upon further doping to ultimately  disappear  with the emergence of  Fermi liquid scattering  when  superconductivity is no longer found at     high enough carrier concentration.\cite{Jin11} The  well defined  Fermi surface at optimal $n$-doping with hot spots of antiferromagnetic scattering    suggests that   mutual reinforcement with d-wave Cooper pairing should also come into play in these materials.\cite{Honerkamp01b} Furthermore  from a high order  RGcalculations, the interplay between cold and hot regions of a two-dimensional Fermi surface is shown to be favorable to unconventional  criticality and NFL behavior  over  extended regions of the FS.\cite{Hartnoll11,Metlitski10}

The connection  between anomalous resistivity and Cooper pairing in cuprates has been further emphasized recently from  the anisotropy displayed  by the  scattering rate, as   extracted from angular dependent magnetoresistance.\cite{AbdelJawad06,AbdelJawad07} Anisotropy of the scattering rate in the copper oxides planes  was found to  simulate  the angular  profile of the SCd gap amplitude.  A RG analysis of the 2D Hubbard model close to half-filling has shown that some mutual influence of electron-hole and electron-electron scattering channels at large momentum transfer can reproduce the d-wave like modulation of the scattering rate.\cite{Ossadnik08,Honerkamp01b}
 
In this respect  the parallel with the  results     obtained here for the $\tau^{-1}_{k_\perp}$ anisotropy   in the quasi-1D case  deserves to be underlined. In   Fig.~\ref{akp} the amplitude of the linear-$T$ scattering was found to mimic  the variation of the d-wave gap modulus on the open FS.  The anisotropy emerges in  the CW domain of regular    spin  fluctuations, namely where pairing interference modifies the momentum dependence of interactions. This shows up in the scattering rate on the FS, which sees the cooling of  nodal    regions concomitant with  the warming of  regions  where the    d-wave gap is maximum.  

Before closing this discussion, it is worth coming back to the limitations of the present approach   when antinesting  is tuned sufficiently close to the critical $t_\perp'^*$ for superconductivity and the scaled scattering amplitudes become  sufficiently large that they invalidate the  quasi-particle picture on the Fermi surface.  Although these strong coupling effects disclose the limitations of  one-loop approximation for the couplings entering in  two-loop self-energy calculations, they may suggest, however,  that  the conditions for a pseudogap formation are combined. Actually  at the approach of $t_\perp'^*$,  the sharp drop of $T_{\rm SDW}$  indicates that the latter should be linked  to a  narrow $t_\perp'$-interval  from $ t_\perp'^*$ where  SDW fluctuations effects are large.  Within this  interval,   the whole Curie-Weiss temperature domain would be characterized by near critical SDW fluctuations likely to destroy quasi-particle excitations. Such a weak coupling RG signature of a pseudogap regime is comparable to  the  one discussed  in the context of the 2D Hubbard model near half-filling. \cite{Katanin04,Rohe05} A possible route to a  more controlled description of  this region of the phase diagram  would be to  push the RG technique  beyond the one-loop level for both self-energy and scattering amplitudes at finite temperature. \cite{Bourbon03,Chitov03,Freire05,Tsuchiizu06a,Tsuchiizu07,Freire08}  Such an extension  is under current investigation.\cite{Sedeki10b}

 In conclusion,   we have used the weak coupling RG method to investigate   the normal phase of   quasi-1D superconductors  near  a   spin-density-wave   instability. Interference between  electron-electron and electron-hole   pairings, which enters in the mechanism of  superconductivity and enhancement of spin fluctuations,  has been found to play  a major role in the  NFL  character of the  normal state.  The tuning of nesting deviations was shown to modify  interference between the two scattering channels  and as a result   the  strength of anomalies in single particle quantities. A  correlation of the superconducting    instability line  with non Fermi liquid  behavior was then found to develop. It  can be seen as an extended    quantum critical region of the phase diagram  that goes beyond the usual  confines   of an   ending magnetic critical point. This mechanism of extended criticality  may be a  key factor  in the evolution of   linear-$T$ resistivity not only in the Bechgaard salts under pressure, but also in other unconventional superconductors such as pnictides and cuprates.

 \acknowledgments
C. B thanks S. Brown,   N. Doiron-Leyraud, Y. Fuseya, D. Jerome, Y. Suzumura, L. Taillefer,  M. Tsuchiizu and A.-M. Tremblay for comments and discussion on several aspects of this work. 
 C. B also expresses  its gratitude to the  the National Science and Engineering Research Council  of Canada (NSERC), the R\'eseau Qu\'ebcois des Mat\'eriaux de Pointe (RQMP) and  the {\it Quantum materials} program of Canadian Institute of Advanced Research (CIFAR) for financial support. Computational resources were provided by the R\'eseau qu\'eb\'ecois de calcul de haute performance (RQCHP) and Compute Canada.

\begin{appendix}
\section{One-particle self- energy\label{RGSelf}}
\begin{figure}
 \includegraphics[width=7.0cm]{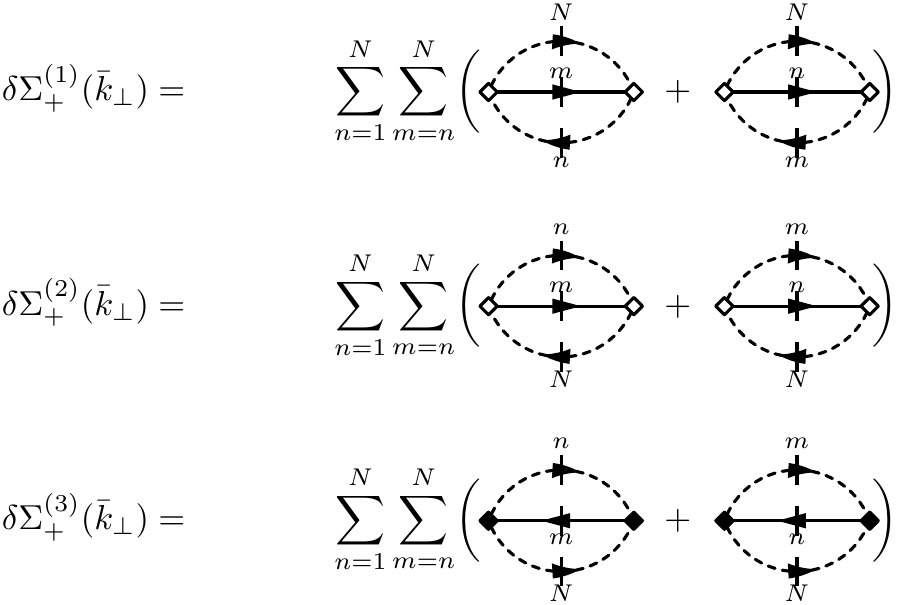}
 \caption{Outer shell contribution to the one-particle self-energy of the $p=+$ branch for normal ($\delta \Sigma_+^{(1,2)}$) and Umklapp ($\delta \Sigma_+^{(3)}$) scattering processes.   $n$ and $m$ refer to the energy indexes in the cascades of contraction, and $N$ to the index of  the running energy shell at $E_N=E_0(\ell)$.\label{self} } 
 \end{figure}

In this appendix, we detail the calculation of the finite temperature outer shell contribution to the one-particle self-energy in the Kadanoff-Wilson scheme of RG. Four our purposes,  it is convenient to discretize the steps of mode elimination by the partial trace operation \cite{Bourbon03,Chitov03}.   Thus after $N \gg 1$ steps of  mode elimination, the scaled energy bandwidth will be noted  $E_N$ ($\equiv E_0(\ell)$). Each
step consists of mode integration within an infinitesimal shell
$E_\Delta$ in the  $E$-space, namely
\begin{equation*}
E_\Delta \equiv \frac{E_0-E_N}{N},
\end{equation*}
where $E_0= 2E_F$; here an equidistant step is chosen for simplicity. After $n$ steps
the energy distance, $E_n$, from the Fermi surface is
\begin{equation*}
\label{Lambdan} E_n={E_0-n\,\,E_\Delta\over 2}, ~~1 \leq n \leq N.
\end{equation*}
We define the $n$-th energy  shell  interval ${\mathcal O}_n$ 
\begin{equation*}
 {\mathcal O}_n \equiv [-E_{n-1}, -E_n] \cup [E_n,
E_{n-1}] \equiv {\mathcal O}_n^- \cup {\mathcal O}_n^+
\end{equation*}
as the union of  shells below and above the Fermi level. 
After the cascade of contractions on the three-particles interactions generated by the RG flow, one obtains the one-particle self-energy corrections of Fig.~\ref{self}.These consist of three types of one-particle self-energy corrections, $  \delta\Sigma_+=\sum_{i=1}^3\delta\Sigma_+^{(i)}$, whose diagrammatic representation of     is shown in Fig.~\ref{self}.
 Here  $\delta\Sigma_+^{(1,2)}$ are associated to the forward and backward 
scattering  contributions and $\delta\Sigma_+^{(3)}$ to Umklapp scattering. 
Their explicit   expressions  are
given by
\begin{widetext}
\begin{equation}
\begin{split}
  \delta\Sigma_+^{(1)}({\bar k}_\perp)=&
              {1\over 2}\left(\frac{\pi\,v_F\,T}{L}\right)^2
              \!\iint \frac{d\,k'_\perp}{2\pi}\,\frac{d\,q_\perp}{2\pi}
                \sum_{\{k,\omega\}}\sum_{n,m} \Big(\!-g^2_2({\widetilde k}_{\perp 1})\
  \!-g^2_1({\widetilde k}_{\perp 1})+g_2({\widetilde k}_{\perp 1})g_1({\widetilde k}_{\perp 1})\Big)\cr
  &\times \,G^0_-(\ktil')\,G^0_-(\ktil'+\qtil)\,G^0_+(\ktil+\qtil)
  \vartheta_N(E_-(\kvec'))\Big(
  \vartheta_m(E_+(\kvec+\qvec))\vartheta_n(E_-(\kvec'+\qvec))+ \{\vartheta_n
    \leftrightarrow\vartheta_m\}\Big)\, ,   
\end{split}
\end{equation}
\begin{equation}
\begin{split}
  \delta\Sigma_+^{(2)}({\bar k}_\perp)=&
              {1\over 2}\left(\frac{\pi\,v_F\,T}{L}\right)^2
              \!\iint \frac{d\,k'_\perp}{2\pi}\,\frac{d\,q_\perp}{2\pi}
                \sum_{\{k,\omega\}}\sum_{n,m} \Big(\!-g^2_2({\widetilde k}_{\perp 2})\
  \!-g^2_1({\widetilde k}_{\perp 2})+g_2({\widetilde k}_{\perp 2})g_1({\widetilde k}_{\perp 2})\Big)\cr
  &\times \, G^0_-(\ktil')\,G^0_-(\ktil'+\qtil)\,G^0_+(\ktil-\qtil)\vartheta_N(E_-(\kvec'))\Big(
  \vartheta_m(E_+(\kvec-\qvec))\vartheta_n(E_-(\kvec'+\qvec))+ \{\vartheta_n
    \leftrightarrow\vartheta_m\}\Big)\, ,
\end{split}
\end{equation}
\begin{equation}
\begin{split}
  \delta\Sigma_+^{(3)}({\bar k}_\perp)=&{1\over 2}\left(\frac{\pi\,v_F\,T}{L}\right)^2
   \!\iint \frac{d\,k'_\perp}{2\pi}\,\frac{d\,q_\perp}{2\pi}
\sum_{\{k,\omega\}}\sum_{n,m} \Big(\!-g^2_3({\widetilde k}_{\perp 3})\
\!-g^2_3({\widetilde k}_{\perp 4})+g_3({\widetilde k}_{\perp 3})g_3({\widetilde k}_{\perp 4})\Big)\cr
  &\times \, G^0_-(\ktil')\,G^0_-(-\ktil'+\qtil)\,G^0_+(-\ktil+\qtil)\vartheta_N(E_-(\kvec'))\Big(
  \vartheta_m(E_+(-\kvec+\qvec))\vartheta_n(E_-(-\kvec'+\qvec))+ \{\vartheta_n
    \leftrightarrow\vartheta_m\}\Big)\, ,   
\end{split}
\end{equation}
\end{widetext}
where   
  $\vartheta_n(E) =1$ for $E \in {\mathcal O}_n$, and $0$ otherwise;  
  $\widetilde{\vartheta}_n(E) = \sum_{m=n}^N \vartheta_m(E)=1$ 
for $ E_N<|E|\leq E_{n}$, and 0 otherwise. The wave vectors are defined as 
  \begin{eqnarray}
  \label{ktildei}
{\widetilde k}_{\perp 1} &=& (k_{\perp}+q_{\perp},k'_{\perp},k_{\perp}),\cr
{\widetilde k}_{\perp 2}&=& (k_{\perp}-q_\perp,k'_{\perp}+q_\perp,k_{\perp}),\cr
{\widetilde k}_{\perp 3}&=& (k_{\perp},-k_{\perp}+q_\perp,k'_{\perp}),\cr
{\widetilde k}_{\perp 4}&=& (k_{\perp},-k_{\perp}+q_\perp,-k'_{\perp}+q_\perp),
\end{eqnarray}

Note that the use of the propagator (\ref{G}) in the calculation of the diagrams of Fig.~\ref{self}  introduces products of the form $g_i\cdot z\cdot z$. In one-loop scheme for interactions,  these $z$ factors  in the self-energy correction are evaluated in the non interacting limit, that is $z=1$. Likewise,   the calculation of self-energy is done with respect to the bare Fermi surface by using the bare propagator in the calculations  for which  all the  $z_\perp$ in (\ref{G})  are fixed to zero.

Following the discretization of the $k_\perp$ interval into $N_p$ 
$(=32)$ patches, such as 
$k'_{\perp},q_{\perp}=-\frac{N_p}{2}\Delta_\perp,..,(\frac{N_p}{2}-1)\Delta_\perp$, 
$\Delta_\perp=\frac{2\pi}{Np}$, the expression becomes
\begin{equation}
\label{sig1}
\begin{split}
  \delta\Sigma_+^{(1)}({\bar k}_\perp)\!=& \frac{1}{2}\iint \frac{d\,k'_\perp}{2\pi}\,\frac{d\,q_\perp}{2\pi}\cr
  & \times  \big(-g^2_2({\widetilde k}_{\perp 1}) - g^2_1({\widetilde k}_{\perp 1})
  +g_2({\widetilde k}_{\perp 1})g_1({\widetilde k}_{\perp 1})\big)\cr
  &\times \Big({\cal I}_1^1({\widetilde k}_\perp,i\omega_\nu)+
  {\cal I}_1^2({\widetilde k}_\perp,i\omega_\nu)\Big).
\end{split}
\end{equation}

The expression for ${\cal I}_1^1$ is given by 
\begin{equation} 
\begin{split}
 & {\cal I}_1^1({\widetilde k}_\perp,i\omega_\nu)\!  =  \!\!\left(\frac{\pi\,v_F\,T}{L}\right)^2 \cr
   & \times \int_{k'_\perp}^{k'_\perp+\Delta_\perp}\!\!\int_{q_\perp}^{q_\perp+\Delta_\perp}\,
  \frac{d\,k''_\perp}{2\pi}\,\frac{d\,q'_\perp}{2\pi}\cr
  &\times \sum_{\{k,\omega\}}\sum_{n,m}G^0_-(\ktil'')
  \,G^0_-(\ktil''+\qtil')\,G^0_+(\ktil+\qtil')\cr
  &\times \vartheta_N(E_-(\kvec''))
  \vartheta_m(E_-(\kvec''+\qvec'))\vartheta_n(E_+(\kvec+\qvec'))
\end{split} 
\end{equation}
 which obeys ${\cal I}_1^2= {\cal I}_1^1|_{\vartheta_n
  \leftrightarrow\vartheta_m}$, and where  
  \begin{equation}
  \label{ktil}
{\widetilde k}_\perp=(k_\perp,k'_\perp,q_\perp).
\end{equation}

Carrying out the frequency summations, we obtain: 
\begin{equation} 
\begin{split}
  &T^2\sum_{\{\omega\}}G^0_-(\ktil'')\,G^0_-(\ktil''+\qtil')\,G^0_+(\ktil+\qtil')\cr
  &=[n_B(E_-(\kvec''+\qvec')-E_-(\kvec''))+n_F(E_+(\kvec+\qvec'))]\cr
  &\times\frac{[n_F(E_-(\kvec''+\qvec'))-n_F(E_-(\kvec''))]}
  {i\omega_\nu+E_-(\kvec''+\qvec')-E_+(\kvec'')-E_+(\kvec+\qvec')}\, ,
\end{split}
\end{equation}
where $n_F(E)$ ($n_B(E)$) is the Fermi (Bose) distribution factor. For a given function 
of energy $f(E)$, we have for the outer-shell integral:  
\begin{equation} 
  \frac{\pi\,v_F}{L}\sum_{{k''}}\vartheta_N(E_-(\kvec''))f(E_-(\kvec''))
  =\frac{\delta E_0(\ell)}{2}\sum_{\mu=\pm}f(\mu\,E_N)\, ,
\end{equation}
which yields 
\begin{equation}
\label{sig1p}
\begin{split}
  \delta\Sigma_+^{(1)}&({\bar k}_\perp)= \frac{1}{2}\iint \frac{d\,k'_\perp}{2\pi}\,\frac{d\,q_\perp}{2\pi}\cr
 &\times \big(-g^2_2({\widetilde k}_{\perp\, 1})-g^2_1({\widetilde k}_{\perp\, 1})
  +g_2({\widetilde k}_{\perp\, 1})g_1({\widetilde k}_{\perp\, 1})\big)\cr
  &\times \Big(-[G^0_+(\ktil)]^{-1}
  \,\,I_1({\widetilde k}_\perp,i\omega_\nu)\,\,d\,\ell
  +I'_1({\widetilde k}_\perp,i\omega_\nu)\,\,d\,\ell
\Big)\, ,
\end{split}
\end{equation}
The expressions for $I_1$ and $I_1'$ are 
\begin{equation} 
\begin{split}
\label{I}
  I_1({\widetilde k}_\perp,i\omega_\nu)&=
  \frac{E_0(\ell)}{4}\sum_{\mu=\pm}
  \iint\frac{d\,k''_\perp}{2\pi}\,\frac{d\,q'_\perp}{2\pi}
  \int_{E_N+|q_{\mu\,1}|/2}^{E_0-|q_{\mu\,1}|/2}d\,q'\cr
  &\left[\mathcal{B}(k''_\perp,q'_\perp,q'-q_{\mu\,1}/2)\!
    +\!\mathcal{B}(k''_\perp,q'_\perp,-q'-q_{\mu\,1}/2)\right]
\end{split}
\end{equation}
and
\begin{equation} 
\begin{split}
\label{Ip}
  I'_1({\widetilde k}_\perp,i\omega_\nu)&=\frac{E_0(\ell)}{4}\sum_{\mu=\pm}
  \iint\frac{d\,k''_\perp}{2\pi}\,\frac{d\,q'_\perp}{2\pi}
  \int_{E_N+|q_{\mu\,1}|/2}^{E_0-|q_{\mu\,1}|/2}d\,q'\cr
  &\left[\mathcal{A}(k''_\perp,q'_\perp,q'-q_{\mu\,1}/2)\!+
    \!\mathcal{A}(k''_\perp,q'_\perp,-q'-q_{\mu\,1}/2)\right],
\end{split}
\end{equation}
and in which we have defined 
\begin{equation} 
\mathcal{A}(k'_\perp,q_\perp,q;q_{\mu\,1})=(2q+\mu\,E(\ell)+q_{\mu\,1})\mathcal{B}(k'_\perp,q_\perp,q;q_{\mu\,1})\, ,
\end{equation} 
and
\begin{equation} 
\begin{split}
  \mathcal{B}(k'_\perp,q_\perp,q;q_{\mu\,1})&=[n_B(-\mu\,E_0(\ell))-q)+n_F(q+q_{\mu})]\cr
  &\times\frac{[n_F(\mu\,E_0(\ell))-n_F(-q)]}{\omega_\nu^2+(2q+\mu\,E_0(\ell)+q_{\mu\,1})^2}\, ,
\end{split}
\end{equation}
 where  $q_{\mu\,1}=\mu\,E_0(\ell)+E_+(\kvec)+A_C(k'_\perp,\qperp)+A_C(\kperp,\qperp)$, and 
$A_C(\kperp,\qperp)=\epsilon_\perp(\kperp+\qperp)-\epsilon_\perp(\kperp)$. 

The outer shell calculation of the remaining two-loop terms $\delta\Sigma_+^{(2,3)}$ are 
obtained in the same way. We only give the final results: 
\begin{equation}
\label{sig2}
\begin{split}
  \delta\Sigma_+^{(2)}&({\bar k}_\perp)= \frac{1}{2}\iint \frac{d\,k'_\perp}{2\pi}\,\frac{d\,q_\perp}{2\pi}\cr
 &\times \big(-g^2_2({\widetilde k}_{\perp\, 2})-g^2_1({\widetilde k}_{\perp\, 2})
  +g_2({\widetilde k}_{\perp\, 2})g_1({\widetilde k}_{\perp\, 2})\big)\cr
  &\times \Big(-[G^0_+(\ktil)]^{-1}
  \,\,I_2({\widetilde k}_\perp,i\omega_\nu)\,\,d\,\ell
  +I'_2({\widetilde k}_\perp,i\omega_\nu)\,\,d\,\ell
\Big)\, ,
\end{split}
\end{equation}
\begin{equation}
\label{sig3}
\begin{split}
    \delta\Sigma_+^{(3)}&({\bar k}_\perp)= \frac{1}{2}\iint \frac{d\,k'_\perp}{2\pi}\,\frac{d\,q_\perp}{2\pi}\cr
 &\times \big(-g^2_3({\widetilde k}_{\perp\, 3})-g^2_3({\widetilde k}_{\perp\, 4})
  +g_3({\widetilde k}_{\perp\, 3})g_3({\widetilde k}_{\perp\, 4})\big)\cr
  &\times \Big(-[G^0_+(\ktil)]^{-1}
  \,\,I_3({\widetilde k}_\perp,i\omega_\nu)\,\,d\,\ell
  +I'_3({\widetilde k}_\perp,i\omega_\nu)\,\,d\,\ell
\Big)\, ,
\end{split}
\end{equation}
where $I_i^{(')}({\widetilde k}_\perp,i\omega_\nu)=
I_1^{(')}({\widetilde k}_\perp,i\omega_\nu)|_{q_{\mu\,1}\rightarrow q_{\mu\,i}}$ ($i=2,3$), with
\begin{equation} 
\begin{split}
\label{I23}
 q_{\mu\,2}&=\mu\,E_0(\ell)-E_+(\kvec)+A_C(k'_\perp,\qperp)-A_C(\kperp,-\qperp)\cr
 q_{\mu\,3}&=\mu\,E_0(\ell)+E_+(\kvec)+A_P(k'_\perp,\qperp)+A_P(\kperp,\qperp)
\end{split}
\end{equation}
where $A_P(\kperp,\qperp)=-\epsilon_\perp(\kperp-\qperp)-\epsilon_\perp(\kperp)$.
Finally, at the step $\ell +d\ell$, the one-particle propagator takes the form :
\begin{equation}
[z(\bar{k}_\perp)G^0_+(\bar{k})]^{-1} + z_\perp(\bar{k}_\perp) -[z(\bar{k}_\perp)]^{-1}\,\delta \Sigma_+(\bar{k})
\label{DG}
\end{equation}
There follows from Eqs. (\ref{sig1p}), (\ref{sig2}), and (\ref{sig3}) the flow equations  (\ref{zeq}) and   (\ref{zpeq}) of Sec.~\ref{Self}.

\end{appendix}

 \bibliography{/Users/cbourbon/Dossiers/articles/Bibliographie/articlesII.bib}

\begin{thebibliography}{93}
\bibitem[{Pow()}]{PowerFit}
\bibinfo{note}{For the same temperature interval, we have checked that the
  power law $\tau^{-1} = c T^\mu$ fits almost just as well the low temperature
  scattering rate than the polynomial two-component expression Eq.~\ref{Poly}.
  In this case, the exponent $\mu$ evolves continuously from linear ($\mu =1$)
  close to $t_\perp'^*$ toward quadratic ($\mu =2$) at large $t_\perp'$ where
  $T_c\to0$.}
\end{thebibliography}
\end{document}